\documentclass[12pt]{JHEP3}
\usepackage{amsmath,amsfonts,amssymb,dsfont,ifthen,epsfig,graphicx}
\usepackage{epsfig,multicol}
\usepackage{epsf}
\usepackage{epstopdf}
\input{epsf.sty}
\newcommand{\be}{\begin{equation}}
\newcommand{\ee}{\end{equation}}
\newcommand{\ba}{\begin{eqnarray}}
\newcommand{\ea}{\end{eqnarray}}
\newcommand{\bas}{\begin{eqnarray*}}
\newcommand{\eas}{\end{eqnarray*}}
\newcommand{\D}{\overline{D}}
\newboolean{Notes}\setboolean{Notes}{true}
\newcommand{\notes}[1]
            {\ifthenelse{\boolean{Notes}}{{\tt #1}}{}}

\def\ap{\alpha^{\prime}}
\def\h{H}
\def\bet{u}
\preprint{hep-th/0611353}

\title{Structures in the Gauge/Gravity Duality Cascade}
\author{Girma Hailu\thanks{hailu@lepp.cornell.edu} and S.-H. Henry
Tye\thanks{tye@lepp.cornell.edu}
\\
Newman Laboratory for Elementary Particle Physics\\
Cornell University \\
Ithaca, NY 14853 }

\abstract{\\ We study corrections to the anomalous mass dimension and
their effects in the Seiberg duality cascade in the
Klebanov-Strassler throat, where $\mathcal{N}=1$ supersymmetric
$SU(N+M)\times SU(N)$ gauge theory with bifundamental chiral
superfields and a quartic tree level superpotential in four
dimensions is dual to type IIB string theory on $AdS_5 \times T^{1,1}$
background. Analyzing the renormalization group flow of the
couplings on the gauge theory side, we propose specific corrections to the
anomalous mass dimension. Applying gauge/gravity duality, we
then show that the corrections reveal structures on the supergravity side
with steps appearing in the running of the fluxes and the metric.
The ``charges" at the steps provide a gravitational source for
Seiberg duality transformations. The finiteness of these corrections suggests
that the theory flows to a baryonic branch rather than to a confining branch.
The cosmological implication of the duality cascade and the gauge/gravity
duality on the brane inflationary scenario and the cosmic microwave background
radiation is pointed out.

}

\begin{document}

\setcounter{equation}{0}
\section{Introduction\label{intr}}

Recent developments in flux compactification in string theory have
provided us with many explicit realizations of the brane world
scenario with stabilized moduli
 \cite{Giddings:2001yu,Kachru:2003aw}. In a typical solution in type
IIB theory, the compactified manifold has a number of warped
throats. It is likely that our standard model particles are open
string modes of a stack of D-branes sitting at the bottom of such a
throat. A prime example of such a throat is the Klebanov-Strassler
(KS) throat \cite{Klebanov:2000hb}; {\it i.e.}, a warped deformed conifold
in type IIB theory on $AdS_5 \times T^{1,1}$ background. Its gauge
theory dual is $\mathcal{N}=1$ supersymmetric $SU(N+M)\times SU(N)$
with bifundamental chiral superfields and a quartic
tree level superpotential which undergoes a cascade of Seiberg
duality transformations \cite{Seiberg:1994pq}. Here, we like to explore the properties of
such a throat in some details. We will start with studying the
running of the couplings in the gauge theory and calculate
corrections to the anomalous mass dimension which dictate the flow
of the gauge theory. The corrections depend on the ranks of the
gauge groups in the duality cascade. We then apply gauge/gravity
duality and find that including the corrections to the anomalous
dimension on the gauge theory side reveals step-like structures in
the metric and the fluxes on the gravity side.

These steps may be observable in cosmology.
Implications of sharp features and/or non-Gaussianity in the cosmic microwave background radiation
due to steps in the inflaton potential have been studied \cite{Adams:2001vc,Chen:2006xj}.
It has become clear that the brane inflationary scenario
in string theory is quite robust \cite{Dvali:1998pa}.
Here, the inflaton is simply the position of the $D$3-brane.
In the simple but realistic KKLMMT scenario, inflation takes place as a
$D$3-brane moves down a warped throat \cite{Kachru:2003sx}.
The duality cascade feature shows up  in the $D3$-brane potential in the warped geometry  as steps.
Such steps in the inflaton potential can introduce sharp features in the cosmic microwave
background radiation, which may have been observed already \cite{Spergel:2006hy,Covi:2006ci}.
So the steps in the throat, though small, can have distinct observable
stringy signatures in the cosmic microwave background radiation.
This point is perhaps best expressed by quoting WMAP \cite{Peiris:2003ff}
: ``a very small fractional change in the inflaton potential amplitude,
$c \sim 0.1 \%$, is sufficient to cause sharp features in the angular power spectrum.''
The possibility of detecting and measuring the
duality cascade is a strong enough motivation to study the throat
more carefully. Although both Seiberg duality and gauge/gravity
duality are strongly believed to be true, neither has been proven;
so a cosmological test is highly desirable. This will also provide strong evidence for string theory.
The steps also show up in the Dirac-Born-Infeld action for brane inflation, which may be observed
separately \cite{Silverstein:2003hf}.

To find the step structure on the gravity side, we shall use a gauge/gravity duality mapping between the couplings in the gauge theory and the dilaton and the backreaction potential in the gravity theory.
On the gauge theory side, as the theory flows
towards the infrared (IR), the larger of the two gauge factors
undergoes a Seiberg duality transition as it becomes strongly
coupled while the weaker factor is treated as a flavor symmetry:
$SU(N+M)$ with $2N$ flavors $\rightarrow SU(2N-(N+M))=SU(N-M)$ with
$2N$ flavors \cite{Seiberg:1994pq}. Repeating such transformation,
the $SU(N+M) \times SU(N)$ gauge theory undergoes a series of
Seiberg duality transitions as it flows towards the IR; {\it i.e.}, the bottom of the
throat. This is the duality cascade. At the $l^{th}$ step
($l=1,2,...$), the gauge theory makes the transition from the
$l^{th}$ region; {\it i.e.}, $SU(N+ M - (l-1)M)\times SU(N-(l-1)M)$ to the
$(l+1)^{th}$ region with $SU(N+M-lM)\times SU(N-lM)$ gauge group.
This duality cascade should lead to steps in the fluxes and the
metric in the gravity side. To see this, let us first look at the
value of the anomalous mass dimension $\gamma$, since the
renormalization group flow of the couplings depends crucially on it.
The $M=0$ case is the conformal Klebanov-Witten (KW) model
 \cite{Klebanov:1998hh}, where $\gamma_{0}= -1/2$. Turning on $M$
breaks the conformal symmetry and so should lead to a correction to
$\gamma$. Intuitively, when $N \gg M$, this correction is expected to
be small and so is usually neglected. Here we find interesting
physics associated with this correction to the anomalous dimension.
Furthermore, this correction becomes substantial as we approach the
IR limit.

Since the gauge theory has the discrete symmetry $M \rightarrow -M$,
$N \rightarrow N+M$, $\gamma$ must be even under this symmetry and
so its leading order correction must have the form $M^{2}/N(N+M)$.
On the other hand, if we turn off the superpotential, the two gauge
couplings will have individual fixed points (when the other group is
weakly coupled and is treated as a flavor symmetry) provided
$\gamma_{N+M} = -1/2 -3M/2N$ and $\gamma_{N} = -1/2 +3M/2(N+M)$,
respectively. So, when the superpotential is turned back on, we
expect the common $\gamma$ to be somewhere in between, and the
renormalization group flow of the couplings take place somewhere in between too.
Since the above symmetry maps $\gamma_{N+M} $ and $\gamma_{N}$ into each other,
we propose that $\gamma$ should be the symmetrization (i.e., average) of $\gamma_{N+M}$
and $\gamma_{N}$.
Similarly, after $l$ steps in the Seiberg duality cascade, the gauge
theory becomes $SU(N+M-lM)\times SU(N-lM)$, with the corresponding
anomalous dimension,
 \ba
 \gamma = -\frac{1}{2} -
\frac{3M^{2}}{4N(N+M)} \rightarrow
 -\frac{1}{2} - \frac{3M^{2}}{4(N-lM)(N+M-lM)}
\label{keyformula}
\ea
Thus $\gamma$ jumps from one value to another value as the
renormalization group flow passes a Seiberg duality
transition. It is this jump in $\gamma$ which causes the steps in
the metric and in the fluxes on the gravity side as one moves towards small $r$, which is roughly the distance away from the bottom of the throat. Note that, towards
the bottom of the throat, $l \rightarrow N/M$, the correction in
$\gamma$ is no longer negligible. For the anomalous dimension to stay finite in the $N=KM$ case,
$l=1, 2, ..., K-1$. That is, there are only $K-1$ steps in the Seiberg duality cascade, and
the infrared flow from $SU(3M)\times SU(2M)$ to  $SU(2M)\times SU(M)$ takes $\gamma=-7/8$.
Of course, higher order corrections may be important when $l$ is large (i.e., when the effective
$K$ is not large).
Note that the correction term in (\ref{keyformula}) blows up for $l=K$.  This divergence leads to an infinite size step. Since higher order corrections have to obey the above discrete symmetry, they will diverge here too. This may signal a breakdown of perturbative corrections to the anomalous dimension in powers of ${M^{2}}/{(N-lM)(N+M-lM)}$. Alternatively, to avoid this divergence, this may signal that the Seiberg duality transition stops at $SU(2M)\times SU(M)$. This latter interpretation means that the theory flows to a baryonic branch rather than to a confining branch.

In the absence of the corrections to the anomalous mass dimension (\ref{keyformula}), the  Seiberg duality cascade is completely smooth when one looks at the geometry in the supergravity side. On the gauge theory side, we see that one of the gauge couplings in the renormalization group flow goes from small to large between duality transformations. Since the anomalous dimension comes in the flow of the dilaton and the backreaction potential and its magnitude changes across Seiberg duality transitions, the warp factor has steps on the supergravity side. Of course, each step should have a width (proportional to the step position), so that steps are actually smoothed out. The resulting warp factor has a cascading behavior that agrees with the picture presented in Ref.\cite{Strassler:2005qs}.

Let us consider the structure in the warp metric $h(r)$ that rescales masses, where
$ds^2=h^{2}(r)d x^{\mu} d x_{\mu} + h^{-2}(r)(dr^2+r^2 ds_{5}^2)$ and $h(r)\sim r$.
The size of the step at $r=r_{p}$ in the warp factor $h(r)$ for large $p < K$ (where $r=r_{K+1}$ is at the edge of the throat) is estimated perturbatively to be (with string coupling $g_{s}$
valued at the step),
\be
\frac{\Delta h(r_{p})}{h(r_{p})} = \frac{ h_{>}(r_{p}) -  h_{<}(r_{p})}{h(r_{p})} \simeq \left(\frac{3g_sM}{2\pi}\right)\frac{1}{p^3}.
\label{key2}
\ee
where $h_{>}(r_{p})$ is the warp factor at $r \ge r_{p}$ and $h_{<}(r_{p})$ that at $r \le r_{p}$.
Although the steps appear as infinitely sharp, this is clearly a consequence of the approximation used here. In reality, we expect the steps to have widths, so the warp factor has a cascading feature. We expect the sharpness of a step to be dictated by the strong interaction scale $\Lambda$
there. So, in general, we expect this multi-step feature to be generic; as we approach the infrared (decreasing $r$), the step size grows (as $p$ decreases) and becomes sharper (as $\Lambda \sim r$ decreases). The spacings between steps are roughly equal as a function of $\ln r $ for large
effective $p$.
These features may show up in brane inflation as signature of the Seiberg duality transition and the gauge/gravity duality in string theory.

Note that the anomalous dimension (\ref{keyformula}) is not the result of a rigorous derivation. We expect that further improvement on the anomalous dimension will introduce a $\Lambda$ dependence. Since the warp factor we have here is not from an explicit solution of the SUGRA equations, and we do expect a correction to the gauge/gravity duality dictionary that is not included here, so the result presented here should be treated as tentative only. Also, the quantitative properties of the warp factor (and so the inflaton potential) may be sensitively dependent on the details of the warped geometry of the throat. On the other hand, this cascading feature should be quite generic even if $V_{D3}$ itself can be quite sensitive to the details of the model.

The rest of this paper is organized as follows. First we give a
brief background review. We then present the details on the
determination of the anomalous mass dimension and its implications on the
renormalization group flow. A study of the implication
in the gravity side is done on the setting of a singular conifold which is
a good approximation to the deformed geometry in the UV region near the edge of the throat. This is good enough for our purpose here as it captures the important features of the physics: the steps with their order of magnitudes and radial locations. We see that the dilaton and the 2-form NS-NS potential run with kinks and the NS-NS flux has steps. This leads to steps in the warp factor. We write down the $D$3-brane world volume action, which is suitable for the study of new features in the KKLMMT inflationary scenario. We comment on the mapping of the flow of the gauge theory to the supergravity flow, suggesting that the mapping/dictionary in the gauge/gravity duality should have corrections as well.
We will then continue with analyzing the supergravity side using $SU(3)$ structures and see how the corrections on the gauge theory side could give rise to geometric obstructions on the supergravity side which provide special locations and sources for Seiberg duality transformations. A full supersymmetric solution on the supergravity side containing the corrections will involve a detailed analysis of the supergravity equations of motion and their solutions and we will not attempt to do that here. We will conclude with some remarks.

\section{Brief review}

Klebanov and Witten found, shortly after
the first example of a dual gauge/gravity theory was given by Maldacena \cite{Maldacena:1998re}, Gubser, Klebanov and Polyakov \cite{Gubser:1998bc}, and Witten \cite{Witten:1998qj}, that type IIB
string theory with a stack of $N$ $D$3-branes on $AdS_5 \times
T^{1,1}$ was dual to $\mathcal{N}=1$ supersymmetric $SU(N)\times
SU(N)$ conformal gauge theory with bifundamental chiral superfields
$A_1$ and $A_2$ transforming as $(\Box,\,\bar{\Box})$ and $B_1$ and
$B_2$ transforming as $(\bar{\Box},\,\Box)$ and a quartic tree level
superpotential \cite{Klebanov:1998hh}.
The quartic tree level superpotential in this theory is given
by
\be
W_{\mathrm{tree}}=w\Bigl((A_1 B_1)(A_2 B_2) - (A_1 B_2)(A_2 B_1)
\Bigr), \label{wtreequartic}
\ee
where color indices from the same gauge group are contracted and $w$ is the tree level coupling.
Let us define the classical dimensionless coupling related to
the tree level coupling $w$ by $\eta=\ln (w/\mu^{1+2\gamma_\eta})$, where $\mu$ has the
dimension of mass. The physical $\beta$ functions of the two gauge
couplings are $\beta_{g} = 3N-2N(1-\gamma_{g})$, and that of $\eta$
is given by $\beta_{\eta}=1+2 \gamma_{\eta}$, where the $\gamma$s
are the anomalous mass dimensions. Although the superpotential
breaks the flavor symmetry to its diagonal version, there is enough
symmetry left so that there is a common anomalous mass dimension,
that is, $\gamma=\gamma_{\eta}=\gamma_{g}$. The theory has a
nontrivial conformal fixed point, where the physical $\beta$
functions of the two gauge couplings associated to the two group
factors in $SU(N)\times SU(N)$ and that of $\eta$ all vanish. This
happens for the same value of the anomalous mass dimension, namely
$\gamma_{0} =-1/2$. So the theory is conformal with this value of
$\gamma$, which is independent of $N$. The stack of $D$3-branes
induces a 5-form R-R flux and the supergravity geometry is a warped
pure $AdS_5 \times T^{1,1}$.

In the KS construction, in a series of papers by Klebanov and
collaborators (with Gubser \cite{Gubser:1998fp}, with
Nekrasov \cite{Klebanov:1999rd}, with Tseytlin \cite{Klebanov:2000nc}
and with Strassler \cite{Klebanov:2000hb}), an additional $M$ number
of $D$5-branes are wrapped near the tip over the $S^2$ cycle of
$T^{1,1}$. See \cite{Herzog:2002ih} for a review.
These wrapped $D$5-branes become fractional $D$3-branes
localized at the apex of the conifold. This enhances the gauge
theory to $SU(N+M) \times SU(N)$ with $A_1,\,A_2\sim
(\Box,\,\bar{\Box})$ and $B_1,\,B_2\sim (\bar{\Box},\,\Box) $. This
theory with the quartic tree level superpotential given by
(\ref{wtreequartic}) is dual to type IIB string theory on a warped
deformed conifold, with $AdS_5 \times T^{1,1}$ background. In this
case, there is no value of common anomalous dimension that makes
the physical $\beta$ functions of the couplings vanish
simultaneously; that is, the addition of the fractional branes makes
the theory nonconformal. The fractional branes induce 3-form R-R flux
through the $S^3$ cycle of $T^{1,1}$. This flux, considered as a
perturbation of the $AdS_5\times T^{1,1}$ background, induces a
2-form backreaction potential which varies with the radius of the
$T^{1,1}$ and produces a logarithmic flow. It was argued that this
theory undergoes a cascade of Seiberg duality transformations with the duality transformation alternating
between the two gauge group factors. The flow of the couplings would
continue until, in the case where $N$ is integral multiple of $M$,
$SU(2M)\times SU(M)$ is left. At this point two different possible routes are discussed in the literature. In one case, a pure $SU(M)$ gauge theory is left in the infrared which undergoes confinement via gaugino condensation \cite{Klebanov:2000hb}. On the string theory side, the confinement corresponds to a deformation of the tip of the cone via geometric transition whereby the $S^2$ cycle is blown-down and the $S^3$ is blown-up with 3-form R-R flux through it. There is also a second possible route for the cascade ending in a baryonic branch of $SU(2M) \times SU(M)$ with a quantum deformed moduli space and a massless axionic moduli field \cite{Gubser:2004qj}. Our work here confirms that the second route is the preferred one.

Assuming that the duality cascade picture is a correct description
of the gauge theory, we start with determining the corrections to
the anomalous dimension. Once we find the corrections to the
anomalous dimension, we will study the effects on the supergravity side
on the setting of the singular conifold. The singular conifold geometry is a good approximate
description in the UV region near the edge of the throat which is enough for our purpose of finding the steps and the corresponding sizes.
We find that the leading corrections in $M/N$
come at orders expected from flux backreaction estimates. For
instance, the leading order correction to the anomalous mass
dimension comes at $\mathcal{O}(M^2/N^2)$, the gauge coupling
$\beta$ functions receive leading $\mathcal{O}(M^2/N)$ corrections
and the dilaton runs at $\mathcal{O}(M^2/N)$.
This is consistent with dual supergravity flux backreaction estimates
 \cite{Klebanov:1999rd,Klebanov:2000nc} where the leading order
corrections to the anomalous dimension is expected to come at most
at $\mathcal{O}(M^2/N^2)$. The magnitude of the
corrections changes after each duality transformation as the matter content of the
theory changes and this
introduces steps in the backreaction $H_3$ flux and in the warp
factor. In $e^{-\Phi}$ for the dilaton and in the $B_2$ NS-NS potential, it is the slope in the logarithmic running which changes
at the cascade steps. The corrections grow as the cascade proceeds
and the difference in the ranks of the gauge groups gets bigger.
Our premise of a changing anomalous dimension as the duality cascade proceeds and the matter content of the theory changes is consistent with the picture of the theory flowing to the baryonic branch with $SU(2M)\times SU(M)$. Here the reason for the flow to a baryonic branch is because an additional Seiberg duality transformation would require an infinite ``charge" at the step.

\setcounter{equation}{0}
\section{Seiberg duality cascade\label{sdccs}}

Let us consider the $\mathcal{N}=1$ supersymmetric $SU(N+M)\times
SU(N)$ gauge theory with chiral superfields transforming as
$A_1,\,A_2\sim (\Box,\,\bar{\Box})$ and $B_1,\,B_2\sim
(\bar{\Box},\,\Box) $ in the KS construction. The
quantity $\gamma=\gamma_A+\gamma_B$ stands for the anomalous
dimension of any one of the objects  $(A_{i}B_{j})$ made out of the
bifundamental chiral superfields, which contains one $A$ and one $B$
superfields and which must have the same anomalous dimension because
of $SU(2)$ global flavor symmetry in the theory. Let us denote the
gauge coupling of the larger group, which is $SU(N+M)$ in the
$1^{st}$ region, by $g_1$, and that of the smaller group, which is
$SU(N)$ in the $1^{st}$ region, by $g_2$, and define $T_1\equiv-2\pi
i \tau_1=8\pi ^2/g_1^2$ and $T_2\equiv-2\pi i \tau_2=8\pi ^2/g_2^2$.
Suppose we start with taking one gauge group as a weakly coupled
gauge theory relative to the other, then we can treat that weaker
group as a flavor symmetry. The running of the physical couplings
 \cite{Shifman:1986zi} with appropriate normalization of the gauge
chiral superfields can then be written as
\be
\beta_{1}=\mu \frac{d T_1(1)}{d\mu}=3(N+M)-2N(1-\gamma_1(1)),
\label{beta1}
\ee
\be
\beta_{2}=\mu \frac{d T_2(1)}{d\mu}=3N-2(N+M)(1-\gamma_2(1)),
\label{beta2}
\ee
and
\be
\beta_{\eta}=\mu \frac{d \eta(1)}{d\mu}=1+2\gamma_{\eta}(1),
\label{betaeta}
\ee
where we have not yet identified the $\gamma$s. We have put
different indices on $\gamma_1(l)$ in (\ref{beta1}), on
$\gamma_2(l)$ in (\ref{beta2}) and on $\gamma_{\eta}(l)$ in
(\ref{betaeta}) since the two gauge groups have different ranks and
``see" different numbers of flavors and would tend to flow with
different anomalous dimensions. The number ``$1$'' in the
parentheses denotes the $l=1^{st}$ region, in the UV region just
before the first duality transformation in the cascade.

According to Seiberg duality, $\mathcal{N}=1$ supersymmetric $SU(N)$
electric gauge theory with $N_f \in (3N/2,3N)$ flavors, which
becomes strongly coupled in the IR, flows to a nontrivial conformal
IR fixed point where it joins a dual $SU(N_f-N)$ magnetic gauge
theory with $N_f$ flavors. Now if we consider the $SU(N+M)$ gauge
theory and think of the other $SU(N)$ gauge group as a weakly gauged
flavor symmetry, we have $\mathcal{N}=1$ supersymmetric $SU(N+M)$
gauge theory with $2N$ flavors; its running is faster than the
running of an $SU(N)$ gauge theory with $2(N+M)$ flavors. Therefore,
the $SU(N+M)$ gauge theory would get strongly coupled faster in the
IR and following Seiberg duality the appropriate description of the
theory in this region is in terms of a weaker dual magnetic theory.
The question of interest to us is the effective value of the
anomalous dimension which dictates the flow. Although it is the
$SU(N+M)$ factor that undergoes duality transformation in the first
step of the cascade, the flow cannot be dictated simply by the fixed
point of $SU(N+M)$ gauge theory with $2N$ flavors for two reasons.
First, the $SU(N)$ group factor which gives a flavor symmetry to
$SU(N+M)$ would itself get strongly coupled during part of the flow.
Second, the running of the tree level coupling has a fixed point for
anomalous dimension $\gamma_{\eta}=-1/2$. In fact, if we consider
the two flows separately, the $SU(N+M)$ factor tends to make
$\gamma<-1/2$ while the $SU(N)$ factor tends to make $\gamma>-1/2$,
and the strengths are slightly different and that is where the
corrections to the anomalous dimension will originate. Consider the
non-trivial IR fixed point of the gauge couplings in the
nonperturbative regime. The anomalous dimension $\gamma_1(1)$ that
would follow from the fixed point of $SU(N+M)$ with $2N$ flavors
($\beta_{1}=0)$ is
\be
\gamma_1(1)=-\frac{1}{2}-\frac{3}{2}\frac{M}{N}.
\label{cascade11}
\ee
Similarly, the anomalous dimension $\gamma_2(1)$ that would follow
from the fixed point of $SU(N)$ with $2(N+M)$ flavors
($\beta_{2}=0)$ is
\be
\gamma_2(1)=-\frac{1}{2}+\frac{3}{2}\frac{M}{N+M}.
\label{cascade12}
\ee
The duality transformation in the first step of the cascade occurs
in the $SU(N+M)$ factor because it would run faster, when the two
gauge factors are looked at separately. In terms of the anomalous
dimensions, it receives more deviation from $-1/2$ than $SU(N)$
does.
The effective value of anomalous mass dimension which guides the running of the
physical couplings should lie somewhere in between.
The gauge theory has the obvious symmetry $M \rightarrow -M$, $N \rightarrow N+M$.
Clearly, $\gamma$ should be even under this symmetry and so cannot
depend on $M/N$ at first order. Note that this symmetry $M \rightarrow -M$, $N \rightarrow N+M$
interchanges $\gamma_1(1)$ (\ref{cascade11}) and $\gamma_2(1)$ (\ref{cascade12}), so that their symmetrization is invariant under this symmetry, i.e.,
\be
\gamma
(1)=-\frac{1}{2}-\frac{3}{4}\frac{M^2}{N(N+M)}.\label{cascs1a}
\ee
That is, $\gamma$ is the average of $\gamma_1$ and $\gamma_2$.
Indeed, we shall see that this leads to results consistent with dual
supergravity flux backreaction estimates
 \cite{Klebanov:1999rd,Klebanov:2000nc} where the leading order
corrections to the anomalous dimension is expected to come at most
at $\mathcal{O}(M^2/N^2)$. Moreover, it gives a picture with the duality cascade ending in a baryonic branch consistent with the discussions in \cite{Gubser:2004qj,Herzog:2002ih,Butti:2004pk}.

Note that one may consider an alternative proposal that still preserves the above symmetry.
Since $\gamma_{\eta}=-1/2$ already yields a vanishing $\beta_{\eta}$,
$\gamma=( \gamma_1 + \gamma_2 + c \gamma_{\eta})/(2+c)$, so that the coefficient in the correction in the anomalous dimension (\ref{cascs1a}) becomes $3/4 \rightarrow 3/2(2+c)$. For generic $c$, this coefficient remains non-zero and positive, so most of the qualitative discussions below still hold.

The fact that correction in (\ref{cascs1a}) comes in the form $M^2/(N(N+M))$ is dictated by the $M \rightarrow -M$, $N \rightarrow N+M$ symmetry and the factor of $-3/4$ comes from the assumption that the effective value of anomalous dimensions which dictates the flow must lie between $\gamma_1$ and $\gamma_2$ and the averaging. Our conclusions including that the gauge theory cascade ends in $SU(2M)\times SU(M)$ depend on the form of $M/N$ combination dictated by symmetry. Corrections of order $\mathcal{O}(M/N)$ to the anomalous dimension were discussed in \cite{Strassler:2005qs}. The corrections to the anomalous dimension we have here come at order $\mathcal{O}(M^2/N^2)$.  Moreover, $\alpha'$ corrections of order $\mathcal{O}(M^4/N^4)$ near the bottom of the throat were
discussed in \cite{Frolov:2001xr}. Our interest here is classical supergravity and the constraints from the gauge/gravity duality are $N \gg 1$, $M \gg 1$, $g_sN \gg 1$ and fixed. One also needs $g_sM$ not to be too small to trust the supergravity side at smaller $r$ where the ranks of the gauge groups become of order $M$. We still want $M$ to be a small perturbation on $N$, $M \ll N$. Our correction is smallest in the early stage of the flow where the rank of the gauge theory is $SU(N+M)\times SU(N)$ and stringy loop corrections would  come at $\mathcal{O}(g_s)\sim 1/N$. Our corrections are bigger than loop corrections so far as $1/N<M^2/N^2$ or $M^2>N$. In the bottom region of the throat, where $N_{\mathrm{eff}}\sim M$ and the 't Hooft coupling is $\sim g_sM$, stringy loop corrections would  be $\mathcal{O}(g_s)\sim 1/M$ and our corrections are bigger than loop corrections, since $1/M< 1$.

The resulting gauge theory after the first duality transformation is
$SU(N-M)\times SU(N)$ and now the running of the $SU(N)$ factor
would be faster and it is its turn for a duality
transformation as the renormalization group flows towards the IR.
Now consider the gauge group
$SU(N-(l-2)M)\times SU(N-(l-1)M)$ in the $l^{th}$ region approaching
the $l^{th}$ step in the duality cascade. For odd $l$,
it is the gauge group whose parent is the gauge factor $SU(N+M)$
which undergoes a duality transformation, while for even $l$ it is
the one with the $SU(N)$ parent. It is convenient to use $g_1$ for
the stronger gauge coupling which corresponds to the gauge group
that undergoes a duality transformation and $g_2$ for the other from
now on. Again treating one gauge group as a flavor
symmetry to the other, we have, for the flow from the $(l-1)^{th}$ to
$l^{th}$ duality transformations,
\be
\gamma_1(l)=-\frac{1}{2}-C_{l-1},
\label{cascs1l}
\ee
\be
\gamma_2(l)=-\frac{1}{2}+C_{l-2},
\label{cascs2l}
\ee
where
\be
C_{l}\equiv \frac{3}{2}\frac{M}{N-l M}. \label{cldef}
\ee
The anomalous dimension for the $l^{th}$ region in the cascade then
follows from the symmetrization of the two,
\be
\gamma(l)=-\frac{1}{2}-\frac{1}{3}C_{l-2}C_{l-1}= -\frac{1}{2} -
\frac{3M^{2}}{4(N+2M-lM)(N+M-lM)}.\label{cascsl}
\ee
Now we are ready to identify the effective values of anomalous dimensions which dictate the flow,
\be\gamma_{1}(l)_{\mathrm{eff}}=\gamma_{2}(l)_{\mathrm{eff}} = \gamma_{\eta}(l)_{\mathrm{eff}} = \gamma (l).
\ee
With this common value of $\gamma (l)$ given by (\ref{cascsl}), (\ref{beta1}), (\ref{beta2}) and
(\ref{betaeta}) become
\be
\mu \frac{d T_1(l)}{d\mu}=3M-C_{l-2}M, \label{beta1l}
\ee
\be
\mu \frac{d T_2(l)}{d\mu}=-3M-C_{l-1}M, \label{beta2l}
\ee
\be
\mu \frac{d \eta(l)}{d\mu}=-\frac{2}{3}C_{l-2}C_{l-1}.
\label{betaetal}
\ee
Thus, the effective running of the couplings makes $g_1$ get
stronger while $g_2$ gets weaker as the theory flows to the IR. In
some region during the flow the two couplings have about equal
strength. The dimensionless tree level coupling $w$ after the
duality transformation goes like the inverse of that before the
transformation and the corresponding $\beta$ function changes sign.
As the magnitude of the anomalous dimension across each step of the
cascade changes because the changing matter content of the cascading
theory, so do the coefficients in the logarithmic running of the
couplings. We see from the $\beta$ functions in (\ref{beta1l}) and
(\ref{beta2l}) that the physical running of the gauge couplings has
appropriate feature with $\mathcal{O}(M^2/N)$ corrections. We will later discuss the
running of the gauge couplings further.

Let us check if the magnitudes of the anomalous mass dimensions we have here
are within range for Seiberg duality. The mass dimension of $\mathcal{O}\equiv(A_1 B_1)(A_2 B_2) - (A_1 B_2)(A_2 B_1)$, $d[\mathcal{O}]$, for the flow from the $l^{th}$ to the $(l+1)^{th}$ duality transition points is
\be
d[\mathcal{O}]=4+2\gamma=
3 -
\frac{3M^{2}}{2(N-lM)(N+M-lM)}.\label{dop-1}
\ee
For the flow involving $\frac{N}{M}-1$ duality transitions, we have
$\frac{9}{4}\le d[\mathcal{O}] \le 3-\frac{3}{2}\frac{M^2}{N(N+M)}$. For the mesons, $\frac{9}{8}\le d[(A_iB_j)] \le \frac{3}{2}-\frac{3}{4}\frac{M^2}{N(N+M)}$. Thus the operator $\mathcal{O}$ is relevant throughout the flow. Note that the mesons have mass dimension $d[(A_iB_j)] \ge \frac{9}{8}$ and this is consistent with the $\ge 1$ bound for the mass dimension of scalars at a conformal fixed point. Here the theory flows nearby such a fixed point. The mass dimension  $d[(A_iB_j)]$ changes from $\frac{11}{8}$ to $\frac{9}{8}$ across the last $(\frac{N}{M}-1)^{th}$ duality transition and a further duality transformation would have made the mass dimension of the mesons $<1$ which would be inconsistent. Here, $d[(A_iB_j)]$ actually diverges for $l=N/M$. To avoid this, the duality cascade should have only $K-1$ transitions and ends with $SU(2M)\times SU(M)$.

\section{Supergravity side}

\subsection{Type IIB supergravity action\label{sugrad}}

Type IIB supergravity is the effective low energy background of type IIB strings.
In this section we want to review and write down a summary of the action and the general equations of motion of type IIB supergravity consisting the fields of interest to us here. The point here is to see the general relations among the fluxes and the metric. At the same time we will see some of the special cases in which the equations reduce and become simpler to deal with directly.

The pair of 16 component spinors of $\mathcal{N}=2$ supersymmetry in ten dimensions have the same chirality in IIB and the corresponding spinor representation can be written as $16 \oplus 16$. The nonperturbative description of strings contains D$p$-branes, which have $p$ spatial and 1 time dimensions. In the IIB case, $p$ is constrained to take on odd numbers. Our interest here is IIB backgrounds in the presence of D3- and D5-branes, and in particular on $AdS_5\times T^{1,1}$ background with the D5-branes wrapping the $S^2$ cycle of $T^{1,1}$ near the tip. The gauge theory dual to this supergravity theory is a nonconformal $\mathcal{N}=1$ supersymmetric $SU(N+M)\times SU(N)$ with bifundamental chiral superfields and a quartic tree level superpotential. The flow of the theory induces a backreaction 2-form NS-NS potential.
The relevant field content of type IIB supergravity
are a dilaton $\Phi$, RR 0-, 2- and 4-forms $C_0$, $C_2$ and $C_4$,
and NS-NS 2-form $B_2$, with corresponding fluxes ${F}_1=d C_0$,
${F}_3=d C_2$, ${F}_5=d C_4$ and ${H}_3=d B_2$
\cite{Polchinski:1998rr}. We also use the same symbols for the
partial derivatives of the fields, $F_1=\partial C_0$, $F_3=\partial
C_2$, $F_5=\partial C_4$ and $H_3=\partial B_2$ as it should be clear
from context which one is meant. We will use normalization in
which the RR flux from a D$p$-brane satisfies,
\begin{equation}
\int_{S^{8-p}}\star F_{p+2}=\frac{2\kappa^2\tau_p N}{g_s}\,, \quad
\quad \tau_p=\frac{1}{\kappa}\sqrt{\pi}(4\pi^2\alpha')^{(3-p)/2} \,,
\quad \quad \kappa=8\pi^{7/2}g_s \alpha'^2\label{fluxnorm}
\end{equation}
where $F_{p+2}$ is $(p+2)$-form flux, $\tau_P$ is the D$p$-brane
tension, $\kappa$ is the gravitational constant in ten dimensions,
$\alpha'$ is the string scale (Regge slope), $g_{s}$ is the string
coupling and $N$ is the number of D$p$-branes.

For the stack of $N$ regular and $M$ fractional $D$3-branes we have
\begin{equation}
\frac{1}{(4\pi^2\alpha')^2}\int_{T^{1,1}}{F}_5=N\,,\quad \quad
\frac{1}{4\pi^2\alpha'}\int_{S^3}{F}_3=M. \label{fluxs53}
\end{equation}
The bosonic part of type IIB classical effective supergravity action is then, in
Einstein frame,
\begin{eqnarray}
S_{10}&=&\frac{1}{2\kappa^2}\int \Bigl( d^{10}x
\sqrt{-G}\Bigl[R-\frac{1}{2}(\partial \Phi)^2-\frac{1}{2}g_s^2
e^{2\Phi}F_1^2 -\frac{1}{12}e^{-\Phi}H_3^2 \nonumber \\
&&-\frac{1}{12}g_s^2
e^{\Phi}\tilde{F}_3^2-\frac{1}{4.5!}g_s^2\tilde{F}_5^2\Bigl]
-\frac{1}{2}g_s^2C_4\wedge {F}_3 \wedge {H}_3\Bigl), \label{s2bsgra}
\end{eqnarray}
where
\begin{equation}
\tilde{{F}}_5\equiv {F}_5+B_2\wedge {F_3}\,,\quad\quad
\tilde{{F}}_3\equiv {F}_3-C_0 {H}_3\,,\label{F53def}
\end{equation}
$G$ is the determinant of the metric in ten dimensions and $R$ is
the Ricci scalar. The 5-form flux is required to satisfy the self
duality constraint
\begin{equation}
\star \tilde{{F}}_5=\tilde{{F}}_5\label{F3sd}
\end{equation}
and we write
\begin{equation}
\tilde{{F}}_5=\mathcal{F}_5+\star \mathcal{F}_5.\label{f5form}
\end{equation}

The corresponding equations of motion are
\begin{eqnarray}
R_{MN}&=&\frac{1}{2}\partial_M \Phi \partial_N \Phi
+\frac{1}{2}g_s^2 e^{2\Phi} \partial_M C_0 \partial_N
C_0+\frac{1}{4}e^{-\Phi}(H_3)_{MOP}(H_3)_{N}^{OP}\nonumber\\
&& +\frac{1}{4}g_s^2
e^{\Phi}(\tilde{F}_3)_{MOP}(\tilde{F}_3)_{N}^{OP}
+\frac{1}{96}g_s^2(\tilde{F}_5)_{MOPQR}(\tilde{F}_5)_{N}^{OPQR}
\nonumber\\
&& -G_{MN}\Bigl(\frac{1}{48}e^{-\Phi}H_3^2
+\frac{1}{48}g_s^2e^{\Phi}\tilde{F}_3^2 +\frac{1}{960}g_s^2
\tilde{F}_5^2\Bigr),\label{dgmn}
\end{eqnarray}
\begin{equation}
d\star d\Phi=g_s^2 e^{2\Phi} {F}_1 \wedge \star {F}_1-\frac{1}{2}
e^{-\Phi}{H}_3\wedge \star {H}_3 +\frac{1}{2}g_s^2
e^{\Phi}\tilde{{F}}_3\wedge \star \tilde{{F}}_3 ,\label{dphi}
\end{equation}
\begin{equation}
d\star(e^{2\Phi} F_1)=- e^{\Phi}{H}_3\wedge \star
\tilde{{F}}_{3},\label{dc0}
\end{equation}
\begin{equation}
d\star(e^\Phi \tilde{{F}}_3)= {F}_5\wedge {H}_3,\label{df3}
\end{equation}
\begin{equation}
d\star(e^{-\Phi}{H}_3-g_s^2C_0 e^{\Phi} \tilde{{F}}_3)=-g_s^2
{F}_5\wedge {F}_3.\label{dh3}
\end{equation}
\begin{equation}
d\tilde{{F}}_5= {H}_3\wedge {F}_3.\label{df5}
\end{equation}
The uppercase indices $M,N,\dots$ above are for the ten dimensional
spacetime coordinates and $G_{MN}$ is the metric. Multiplying both
sides of (\ref{dgmn}) by $G^{MN}$ gives
\begin{equation}
R=\frac{1}{2}(\partial \Phi)^2+\frac{1}{2}g_s^2 e^{2\Phi}(\partial
C_0)^2 +\frac{1}{24}e^{-\Phi}H_3^2 +\frac{1}{24}g_s^2
e^{\Phi}\tilde{F}_3^2. \label{traceR}
\end{equation}

We note a few general features of the theory from the above equations of motion. From (\ref{dc0}) we see that the $H_3$ and the $F_3$ fluxes are perpendicular when $C_0=0$, which is the case in KS and we will set $C_0=0$ in our analysis from now on unless when we explicitly state otherwise. From (\ref{dphi}) we see that the dilaton would be constant for a precise matching of the $H_3$ and $F_3$ fluxes such that $
e^{-\Phi}{H}_3\wedge \star {H}_3 = g_s^2 e^{\Phi}F_3\wedge \star F_3 $.
In the case when the dilaton is taken constant, the equations of motion simplify and the solution on a singular conifold was found by Klebanov and Tseytlin \cite{Klebanov:2000nc}, with the singularity at tip of the conifold where the radius of $AdS_5$ (or $T^{1,1}$) vanishes. However, with the KS picture in terms of Seiberg duality cascade, confinement via gaugino condensation at the end of the cascade on the gauge theory side leads to a deformed conifold with the tip being $S^3$. In this case, the tip of the cone is smoothed out and cut off at some finite $r$ whose size depends of the magnitude of the 't Hooft coupling in the confining gauge group, $g_s M$, which is related to the glueball superfield \cite{Dijkgraaf:2002fc} with expectation value related to the scale of the confining gauge theory. Thus one needs a metric and flux ansatz which takes into account the interpolation between $S^3$ at the tip and the asymptotically $S^2\times S^3$ geometry at large $r$. With a metric ansatz, one computes the Ricci scalar, and then equate it to (\ref{traceR}) to determine the geometry and obtain the Klebanov-Tseytlin solution \cite{Klebanov:2000nc}. A second special case is where the NS-NS flux $H_3$ is turned on by wrapping NS5-branes on $S^2$ while both R-R fluxes $F_5$ and $F_3$ vanish and the equations are simplified. Now we cannot have a constant dilaton, since the right hand side of (\ref{dphi}) is nonzero. The solution to this case with $\mathcal{N}=1$ supersymmetry was obtained by Maldacena and Nunez (MN) \cite{Maldacena:2000yy, Chamseddine:1997nm, Chamseddine:1997mc}. The KS solution on the deformed conifold with $F_5$ and $F_3$ fluxes turned on and the MN solution with the $H_3$ flux turned on are two well-known regular solutions on type IIB background with $\mathcal{N}=1$ supersymmetry.  A natural question was whether there existed a flow between them. The possibility for this was analyzed and a metric and a flux ansatz for it given by Papadopoulos and Tseytlin \cite{Papadopoulos:2000gj}. This issue was further investigated by Gubser, Herzog and Klebanov \cite{Gubser:2004qj} who found a leading order perturbative solution around the KS solution. Butti, Grana, Minasian, Petrini, and Zaffaroni used $SU(3)$ structures to find a one parameter set of numerical solutions which flow in a direction from KS to MN \cite{Butti:2004pk}.

\subsection{Mapping gauge coupling running to supergravity flow\label{sugrad2}}

In this section we want to apply the
gauge/gravity duality to map the renormalization group flow of the
gauge couplings to the running of the dilaton and the backreaction
NS-NS 2-form potential. We discuss how, with the inclusion of the corrections,
the gauge couplings in their renormalization flow may stay finite throughout the duality cascade.

The stack of $M$ $D$5-branes wrapping $S^2$ of the $AdS_5 \times
T^{1,1}$ background creates 3-form flux through $S^3$ which induces
a backreaction 2-form potential $B_2$ in the $S^2$ cycle. The sum of
the two gauge coupling coefficients $T_+\equiv T_1+T_2$, which can
be taken as the effective gauge coupling, is related to the
effective string coupling containing the dilaton in the dual gravity
theory. The difference between the two gauge coupling coefficients
$T_-\equiv T_1-T_2$ is nonzero because the ranks of the two gauge
groups are different and it describes the nonconformal nature of the
theory. Indeed, $T_-$ must dictate the flows in both the gauge and
the gravity theories. We note that the
supergravity equation of motion in
the presence of nonzero R-R $F_3$ and $F_5$ fluxes from the $D$3- and
$D5$- branes could have a consistent set of solutions only if the
NS-NS 2-form potential $B_2$ is nonzero. Indeed, the two parameters
$T_+$ and $T_-$ on the gauge theory side are mapped to the effective
string coupling and the 2-form potential $B_2$ through the relations
 \cite{Klebanov:1999rd,Morrison:1998cs},
\be
T_1+T_2 =  \frac{2\pi}{g_s e^\Phi} , \label{t1p2phi}
\ee
\be
 T_1-T_2 =  \frac{2\pi}{g_s e^\Phi}({\hat{ b}-1})
 =\frac{2\pi}{g_s e^\Phi}({\bar {b}_2(\mathrm{mod}\,2))}\label{t1m2phi}
  \ee
where
 \be
  \qquad  \hat b-1 = \bar{b}_{2}  \, (\mathrm{mod}\, 2),\qquad \bar{b}_2 =  b_2 -1,
  \ee
and
\be
 b_2\equiv  \frac{1}{2\pi^2 \alpha'}\int_{S^2}B_2. \label{b2def}
\ee

The variable $\bar{b}_2$ measures the deviation of $T_{-}$ from zero due to the backreaction $B_2$ potential from where the two gauge couplings have equal magnitude in $r_{1} < r < r_{0}$. Therefore, the flow is such that $\bar{b}_2=0$ at the first point where the two couplings are equal.
We will arrange the supergravity flow such that $b_2(r)$ vanishes at the edge of the throat, $r=r_0$, and $0 \le  \hat b  \le 2$.
Following from the quantization condition on $H_{3}$, $\pi b_{2}$ must be a periodic variable with period $2 \pi$. This periodicity is crucial for the cascade phenomenon. Note that $b_{2}$ decreases as the theory flows towards the infrared (smaller $r$) and Seiberg duality takes place when $ {\hat b} -1$ reaches $-1$.
For $r_{1} \le r < r_{0}$, we label the couplings as $T_{i}(1)$, where $``1"$ labels the fact that the flow is in the first region, {\it i.e.}, the couplings are for the $SU(N+M) \times SU(N)$ gauge theory.
We see from (\ref{beta1l}) and (\ref{beta2l}) that the running of $T_{1}(1)$ and $T_{2}(1)$,
\be
\mu \frac{\partial T_{1}(1)}{\partial \mu} = 3M -\frac{3}{2} \frac{M^{2}}{N+M}\,, \label{minusflow21}\ee
\be\mu \frac{\partial T_{2}(1)}{\partial \mu} = - 3M -\frac{3}{2} \frac{M^{2}}{N}\,, \label{minusflow22}\ee
\be\mu \frac{\partial T_{-}(1)}{\partial \mu} = 6M +\frac{3}{2} \frac{M^{3}}{N(N+M)}\,.\label{minusflow}
\ee
\begin{figure}[t]
\begin{center}
\leavevmode
\includegraphics[width=1\textwidth, angle=0]{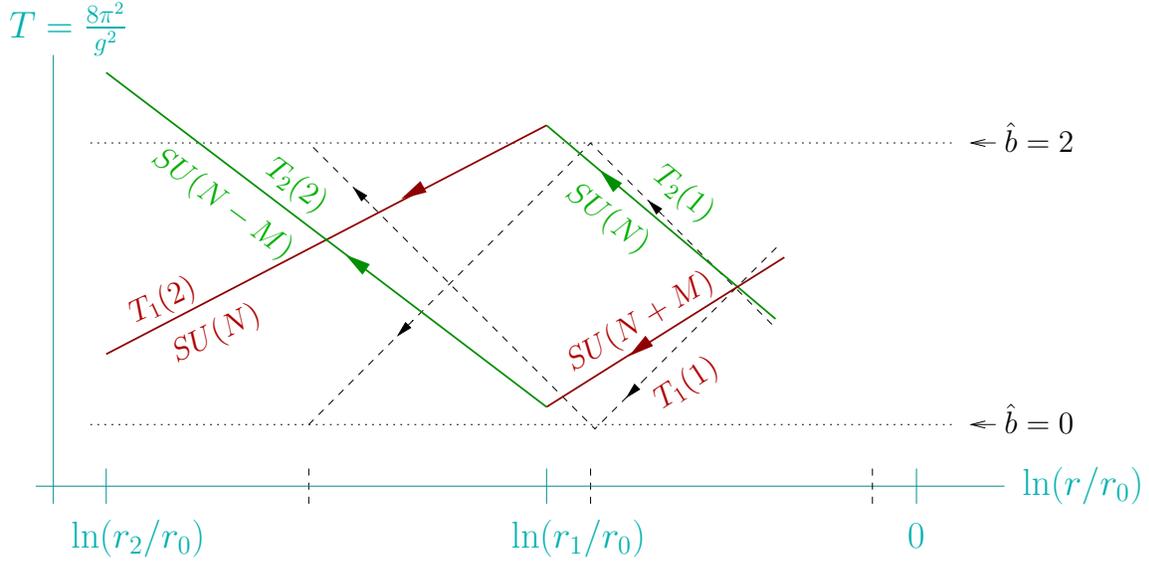}
\caption{A schematic comparison of the flows of the couplings in $SU(N+M) \times SU(N)$ without the corrections (dashed lines) versus that with the corrections to the anomalous mass dimension (solid lines).
Note that a Seiberg duality transition occurs when the flow of $T_{-}(1) =T_{1}(1) -T_{2}(1)$ reaches a period in $b_{2}$.}
\label{sdflow1}
\end{center}
\end{figure}

The flows are illustrated in Figure \ref{sdflow1}.
Let us first consider the KS case without the corrections to the anomalous mass dimension.
Let us start at the value of $r < r_{0}$ where ${\hat b}=1$, $\bar{b}_2= 0$ or $b_{2}=-1$, so that the two gauge couplings
are equal.
The dilaton $\Phi=0$ in KS and the running of $T_{1,2}(1)$ is given by (\ref{minusflow21}) and (\ref{minusflow22}), with $\pm 3M$. As $r$ decreases to $r=r_{1}$,
\be T_{1} \rightarrow 0,  \quad \quad  {\hat b} \rightarrow 0,\ee
that is, $g_{1} \rightarrow \infty$ when $b_{2} \rightarrow -2$. With our notation, $\hat{b}$ takes values between 0 and 2.
At $r=r_{1}$, Seiberg duality transformation occurs, so $T_{1}(1) \rightarrow T_{2}(2)$ and $T_{2}(1) \rightarrow T_{1}(2)$.  Now the flows correspond to that of the gauge theory $SU(N) \times SU(N-M)$, where $T_{1}(2)$ starts decreasing (towards small $r$), that is, $SU(N)$ is getting strongly interacting. Seiberg duality transformation takes place when the strongly interacting coupling $g_{1}$ is infinite. On the other hand, the warped geometry in the supergravity side is completely smooth as we pass through such a duality transition.

Now let us consider the case with the corrections included, with the flows of the couplings given by
(\ref{minusflow}) and schematically illustrated in Figure \ref{sdflow1} as solid lines. We see that $T_{1}(1)$ is decreasing slower while $T_{2}(1)$ is increasing faster as $r$ decreases. Note also that (\ref{minusflow}) shows that $T_{-}(1)$ is flowing faster so, starting at $r$ where the two couplings are equal in $r_{1} < r < r_{0}$, $b_{2}$ reaches a period with the flow tilted with respect to the case without the corrections as shown in Figure \ref{sdflow1}, since the magnitude of the slope in the running of $T_2$ is greater than $T_1$.
The locations of duality transitions happen to occur further away at smaller $r$ with the running of both $\Phi$ and $b_2$ is taken into account in the gauge/gravity duality mapping as shown in Figure \ref{sdflow1}. We  will see that from the values of $r_l$ in Section \ref{sugraflow}. The first duality transition occurs when the effective number of $D3$-branes drops such that the stronger gauge group changes from rank $N+M$ to rank $N$. This translates to a decrease in $b_2$ by 2 units, which takes place at small but finite $T_1$.
For the flow from the $(l-1)^{th}$ to the $l^{th}$ cascade steps we have (\ref{beta1l}) and (\ref{beta2l}),
\be
\frac{d}{d\ln (\Lambda/\Lambda_c)}{T}_+=-
\Bigl(C_{l-2}+C_{l-1}\Bigr)M, \label{t1pt2}
\ee
\be
\frac{d}{d\ln
(\Lambda/\Lambda_c)}{T}_-=\Bigl(6+C_{l-1}-C_{l-2}\Bigr)M,
\label{t1mt2}
\ee
where $\Lambda$ is the scale of the gauge theory and $\Lambda_c$ is
the cutoff. The scale of the gauge theory is mapped to the radial
coordinate $r$ of $AdS_5$ in the dual gravity theory, so we have
\be
\Lambda \sim r. \label{lambr}
\ee
We note that the magnitude of the correction to the running of $T_{-}$
increases with increasing $l$, since $C_{l-1}>C_{l-2}$. This
increase in the coefficient of the logarithmic running of
$T_{-}$ will lead to a change of the slope in the logarithmic
running of the $B_2$ potential which results in a step in the 3-form
NS-NS flux $H_3$ and in the warp factor at a duality transformation.
Applying the derivative with respect to $\ln(r/r_0)$ on (\ref{t1p2phi}) and
(\ref{t1m2phi}), we obtain
\be
\frac{d}{d\ln {(r/r_0)}} \,e^{-\Phi}=-S_{l},\label{t1p2phie}
\ee
\be
\frac{d}{d\ln {(r/r_0)}} \Bigl(e^{-\Phi} {\bar {b}_2 } \Bigr) = D_l,
\label{t1m2phie}
\ee
where $r_0$ is the $AdS_5$ radius at the edge of the throat and we have introduced two sets of
constants,
\be
S_{l}\equiv (C_{l-1}+C_{l-2})\frac{g_s M}{2 \pi},\quad {} D_l\equiv
(6+C_{l-1}-C_{l-2})\frac{g_s M}{2 \pi}.\label{DlSl}
\ee
Notice that $S_{1} \sim 1/K$ while the correction in $D_{1}$ goes as $1/K^{2}$.
Suppose the $l^{th}$ duality transformation takes place at $r=r_l$.
First we can solve (\ref{t1p2phie}) for $\Phi(r)$ in the range $r \ge r_{1}$,
\be
e^{-\Phi(r)}= e^{-\Phi_0} - S_{1}\ln(r/r_{s})= 1- S_{1}[\ln(r/r_{0}) -c_{1}]
 \label{phitsol1}
\ee
where $\Phi_{0}$ may be absorbed into the definition of $g_{s}$ and $c_{1}$ should be determined by an appropriate boundary condition. Note that we have chosen $\Phi(r_s)=\Phi_0$ at some $r=r_s$ and $e^{-\Phi(r_0)}=1+S_1 c_1$. We then have the solution for $\Phi$ in the range $r_{l}\le r\le r_{l-1}$ between the  $(l-1)^{th}$ and the $l^{th}$
duality transformation locations,
\be
e^{-\Phi(r)}= 1 +S_1  c_1-\sum_{k=1}^{l-1}S_{k}\ln(r_k/r_{k-1})
-S_{l}\ln(r/r_{l-1}), \label{phitsol}
\ee
and $\Lambda/\Lambda_c=r/r_0$. The values of $r_l$ will be computed
using the change in magnitude of $b_2$.
The leading term in the
variation of $e^{-\Phi}/g_s$ with respect to $\ln(r/r_{l-1})$ comes at $\mathcal{O}(M^2/N)$ consistent
with flux backreaction expectations \cite{Klebanov:1999rd}. Note that near the bottom of the throat at $r=r_{K-1}$, using $\ln(r_k/r_{k-1})\sim -1/g_sM$ which we will see later, we have $e^{-\Phi(r_{K-1})}-e^{-\Phi(r_{0})}=-\sum_{k=1}^{K-1}S_{k}\ln(r_k/r_{k-1})\sim -(g_sM^2/N)(-1/g_sM)$ $(N/M)=\mathcal{O}(1)$ and, therefore, the change in $e^{-\Phi}$ from the edge to the bottom of the throat is $\mathcal{O}(1)$.

Our interest is first to show how kinks appear in
the running of the dilaton and the $B_2$ potential which lead to
steps in the $H_3$ flux and in the warp factor. The expressions we present
are only approximate and good for the large $r$ region. We will not attempt to find the full supersymmetric
solution on the supergravity side on the deformed/resolved conifold here. We seek a UV approximate expression for $\bar{b}_2(r)$ with $\bar{b}_2(r_0)=1$ (or ${b}_2(r_0)=0$) at the edge of
the throat. This corresponds to the case in which the gauge coupling
$g_1$ just starts getting stronger while $g_2$ starts getting weaker
as the theory starts flowing down from the edge of the throat. We
then obtain from (\ref{t1m2phie}) for $r$ in the range $r_{l}\le
r\le r_{l-1}$,
\be
e^{-\Phi(r)} {\bar{ b}_2}(r)\approx e^{-\Phi(r_{l-1})} {\bar b}_2(r_{l-1})
+D_{l}\ln(r/r_{l-1}). \label{b2sol}
\ee
The $l^{th}$ duality transformation will occur at $r=r_l$ such that
$\bar{b}_2(r_l)=-(2l-1)$. With this and expressing $\ln(r/r_{l-1})$ in terms
of $\Phi$, (\ref{b2sol}) gives
\be
e^{-\Phi(r)} {\bar{b}}_2(r)\approx - (2l-3)e^{-\Phi(r_{l-1})}
+\frac{D_{l}}{S_l}\left( e^{-\Phi(r_0)}
-\sum_{k=1}^{l-1}S_{k}\ln(r_k/r_{k-1})-e^{-\Phi(r)} \right).
\label{b2solphi}
\ee
The 2-form potential for the same range of $r$ then follows from
(\ref{b2def}) and (\ref{b2sol}),
\be
B_2(r)\approx {b}_2(r)\frac{\pi \alpha'\omega_2}{2}, \label{B2sol}
\ee
where $\omega_2$ denotes the $S^2$ cycle in $T^{1,1}=S^2\times S^3$.
The corresponding 3-form NS-NS flux $H_3=dB_2$ is, noting that $\partial b_2/\partial r=\partial \bar{b}_2/\partial r$,
\be
 e^{-\Phi(r)}H_3(r)\approx\Bigl(D_l+ S_l \bar{b}_2(r) \Bigr)\frac{\pi \alpha'
dr \wedge\omega_2}{2r}, \label{H3sol}
\ee
which has steps at Seiberg duality transformation locations, since the magnitudes of $D_l$ and $S_l$ change across locations of duality transitions.
Each time $r$ decreases past $r_{l}$, $\bar{b}_2$  drops by 2, as implied
by Seiberg duality and (\ref{t1m2phi}). Note also that
the steps in the flux give steps in the metric as the
two are related by equations such as (\ref{traceR}). We will study the steps in the warp factor in Section \ref{sugraflow}.

Finally we like to point out that the dictionary in the mapping between couplings in the gauge and the gravity variables (\ref{t1p2phi}) and  (\ref{t1m2phi}) should have corrections. Let us discuss this point in some detail here. First consider the above mapping by introducing the variables
\be
t_+ = t_{1} + t_{2}= \frac{g_s e^\Phi}{2\pi}(T_1+T_2)=1 , \label{ts1p2phi}
\ee
\be
t_- = t_{1} - t_{2}= \frac{g_s e^\Phi}{2\pi}(T_1-T_2)=   \hat{b} -1
\label{ts1m2phi}
\ee
Since
\be
t_1 =\frac{{\hat{b}}}{2}, \quad \quad  t_2 =\frac{2-\hat{b}}{2}
\ee
and $t_{1}, t_{2} \ge 0$, we see that $0 \le {\hat b} \le 2$ and $1  \ge t_{1}, t_{2} \ge 0$.
(This is illustrated in Figure  \ref{sdflow1}.) Let us ignore corrections and start at the value of $r < r_{0}$ where ${\hat b} =1$ so that the two gauge couplings are equal.
As $r$ decreases to $r=r_{1}$,
\be t_1 \rightarrow 0  \quad \quad  {\hat b} \rightarrow 0,\ee
that is, $g_{1} \rightarrow \infty$.
In the KS case, the geometry is smooth as we cross the Seiberg duality transition point. With the corrections coming from the anomalous mass dimension we have here, the steps in the warp factor and the kinks in the dilaton introduces discontinuities. Clearly, we expect such discontinuities to be smoothed out by further correction terms.
That is, we expect that the relations (\ref{t1p2phi}) and (\ref{t1m2phi}) should be modified by corrections.
For example, if the mapping (\ref{t1p2phi}) is not exact, then we expect either a correction on the
gravity side, or equivalently, a correction on the gauge theory side. If the gravity side in  (\ref{t1p2phi}) has a positive correction, either coming from $\ap$ or quantum corrections, such a correction will allow the flow of $g_{1}$ to a large but finite value, as expected. We also expect the anomalous dimension to depend on the scale $\Lambda \sim r$.
Such corrections should smooth out the discontinuities in the warped geometry and the dilaton flow.

\section{Warp factor with steps\label{sugraflow}}

We shall consider the UV region where the correction to the anomalous dimension is relatively small.
In this region, the effect of the deformation of the conifold is small, so we can use the singular conifold geometry.


First we want to summarize the singular conifold metric for the sake of completeness and clarification of our notation. The ten dimensional metric which describes the singular $AdS_5\times
T^{1,1}$ geometry has the form
\begin{equation}
ds^2={\h}^{-1/2}(r)\eta_{\mu \nu} dx^\mu dx^\nu + {\h}^{1/2}(r)(dr^2+r^2
ds_{T^{1,1}}^2),\label{10dmetric}
\end{equation}
where $ds_{T^{1,1}}^2$ is the metric on the $T^{1,1}=S^3\times S^2$
base of the conifold which is
parameterized by five angles $\theta_1,\, \theta_2\,\in [0,\pi]$,
$\phi_1,\, \phi_2\,\in [0,2\pi]$, and $\psi\in[0,4\pi]$. A compact
notation of the metric and the cycles in $T^{1,1}$ is obtained by
introducing the 1-forms \cite{Minasian:1999tt, Klebanov:2000hb},
\begin{eqnarray}
&g^1=\frac{1}{\sqrt{2}}(e^1-e^3)\,,\quad
g^2=\frac{1}{\sqrt{2}}(e^2-e^4),&\nonumber\\
&g^3=\frac{1}{\sqrt{2}}(e^1+e^3)\,,\quad
g^4=\frac{1}{\sqrt{2}}(e^2+e^4)\,,\quad g^5=e^5,&\label{gbasis}
\end{eqnarray}
where
\begin{eqnarray}
e^1=-\sin\theta_1 d\phi_1\,,\quad e^2=d\theta_1\,,\quad
e^3=\cos\psi\sin\theta_2 d\phi_2-\sin\psi d\theta_2\,,\nonumber\\
e^4=\sin\psi\sin\theta_2 d\phi_2+\cos\psi d\theta_2\,,\quad
e^5=d\psi+\cos\theta_1 d\phi_1+\cos\theta_2 d\phi_2. \label{ebasis}
\end{eqnarray}
The metric on $T^{1,1}$ can then be written as
\begin{eqnarray}
ds_{T^{1,1}}^2&=&\frac{1}{6}\sum_{i=1}^4(g^i)^2+\frac{1}{9}(g^5)^2\nonumber\\
&=&\frac{1}{9}\left(d\psi+\sum_{i=1}^{2}\cos \theta_i
d\phi_i\right)^2+\frac{1}{6}\sum_{i=1}^{2}\left(d\theta_i^2 + \sin^2
\theta_i d\phi_i^2\right) \label{metrict11}
\end{eqnarray}
The $S^2$ and $S^3$ cycles of $T^{1,1}$ are represented by the
following 2- and 3-forms
\begin{equation}
\omega_2=\frac{1}{2}(g^1\wedge g^2+g^3\wedge g^4)\,,\quad \quad
\omega_3=\frac{1}{2}g^5\wedge(g^1\wedge g^2+g^3\wedge
g^4)\label{w2w3}
\end{equation}
which give
\begin{equation}
\int_{S^3}\omega_3=8\pi^2\,,\quad\quad \int_{S^2}\omega_2=4\pi.
\label{w3w2per}
\end{equation}
Our notation is such that the inner product of a $p$-form
$\omega_p=\frac{1}{p!}(\omega_p)_{M_1 \dots M_p}dx^{M_1}\wedge \dots
\wedge dx^{M_p}$ satisfies
\begin{equation}
\omega_p \wedge \star \omega_p= \frac{1}{p!}\, (\omega_p)_{M_1 \dots
M_p} (\omega_p)^{M_1 \dots M_p}\, \mathrm{vol}= \frac{1}{p!}\,
\omega_p^2\, \mathrm{vol}.\label{wpinner}
\end{equation}
where vol stands for the ten dimensional volume element which we
write as
\begin{eqnarray}
\mathrm{vol}&=&\frac{r^5\sin\theta_1\sin\theta_2\sqrt{\h}}{108}\,dx^0\wedge
dx^1\wedge dx^2 \wedge dx^3 \wedge dr \wedge d\psi \wedge d\theta_1
\wedge d\theta_2 \wedge d\phi_1  \wedge d\phi_2\nonumber\\
&=&\frac{r^5 \sqrt{\h}}{108}\,dx^0\wedge dx^1\wedge dx^2 \wedge dx^3
\wedge dr \wedge g^1 \wedge g^2 \wedge g^3 \wedge g^4 \wedge g^5.
\label{vol10d}
\end{eqnarray}
Moreover, the above ten dimensional metric gives the Ricci scalar
\begin{equation}
R=-\frac{1}{2 {\h}^{3/2}}\Bigl({\h}''+\frac{5}{r}{\h}'\Bigr).\label{ricci106}
\end{equation}

As we have seen, the corrections lead to a running of the dilaton
with the slope in the logarithmic running of $e^{-\Phi}$ changing
across each cascade step. Moreover, the $H_3$ flux has a step-wise
jump at each cascade step. In this section we want to find the
resulting step-wise corrections to the warp factor.

First let us recall the Klebanov-Tseytlin (KT) solution \cite{Klebanov:2000nc} on the singular conifold. The $N$ regular
$D$3-branes induce flux through $T^{1,1}$, the $M$ fractional
$D$3-branes induce flux through $S^3$, the backreaction $H_3$ flux
is through $dr\wedge \omega_2$. With the quantization given in
(\ref{fluxs53}),
\begin{equation}
{F}_3=\frac{1}{2}M\alpha' \omega_3, \label{f3sol}
\end{equation}
\begin{equation}
{H}_3=\frac{3}{2r}g_s M\alpha' dr\wedge \omega_2,\label{h3sol}
\end{equation}
\begin{equation}
\mathcal{F}_5= \frac{1}{2} \pi \alpha'^2(N+\frac{3}{2\pi} g_s
M^2\ln(r/r_0))\omega_2 \wedge \omega_3.\label{f5sol}
\end{equation}
In this case,
\begin{equation}
H_3^2=g_s^2 F_3^2=\frac{243\alpha'^2 g_s^2
M^2}{{\h}^{3/2}r^6}.\label{h3f3rel}
\end{equation}
The relation between $F_3$ and $H_3$ as given by (\ref{h3f3rel}) is
valid only when the dilaton is constant as we can see from one of
the supergravity equations of motion given by (\ref{dphi}). One then
puts (\ref{ricci106}) and (\ref{h3f3rel}) in (\ref{traceR}), with
$\Phi$ set to zero, and solves for $\h(r)$ to
obtain the KT solution \cite{Klebanov:2000nc},
\be
H_{0}(r) =  \frac{27 \pi {\ap}^{2} g_{s}}{4 r^{4}} \left( N + \frac{3g_{s}M^{2}}{2 \pi} [\ln (r/r_{0}) +1/4] \right),
\label{warpH0}
\ee
where, with the volume of $T^{1,1}$ given by $v \pi^{3}=16\pi^{3}/27$,
\be
 r_{0}^{4} = 4 \pi \alpha'^{2}g_{s}N/v = \frac{27 \pi \alpha'^{2}g_{s}N}{4}
\ee
and the locations of duality transitions in the KS throat are given by
\be
r_{l}= r_{0} \exp\left(-\frac{2l\pi}{3g_s M}\right).\label{rlr0fa}
\ee
Note that the constant piece $N$ in (\ref{warpH0}) is determined by the boundary condition.
We see that the effective $D3$-brane charge is given by
\be
N_{\mathrm{eff}} = N + \frac{3g_{s}M^{2}}{2 \pi} \ln (r/r_{0})
\ee
so that at $r=r_{l}$, $N_{\mathrm{eff}} = N - lM$. The term with $1/4$ factor is introduced to ensure that
the warp factor $H_{0}(r)$ is monotonic for $r_{0} \ge r \ge r_{K}$.

Now let us proceed with the corrections included. First we want to
find the locations $r_l$ where the duality transformations take
place. The relation between $r_{l}$ and $r_{l-1}$ is obtained
remembering that $\bar{b}_2(r_l)-\bar{b}_2(r_{l-1})=-2$ and $\bar{b}_2(r_l)=-(2l-1)$ which
with (\ref{b2sol}) or (\ref{b2solphi}) gives the recursion relation
\ba
\ln\left(\frac{r_l}{r_{l-1}}\right)& = &-\frac{1}{D_l}\Bigl((2l-1)
e^{-\Phi(r_l)}-(2l-3)e^{-\Phi(r_{l-1})}\Bigr)\nonumber\\ &=& -\frac{2}{D_l+
S_l\bar{b}_2(r_l)}e^{-\Phi(r_{l-1})}=-\frac{2}{D_l-(2l-1)
S_l}e^{-\Phi(r_{l-1})}, \label{rrel}
\ea
which gives
\begin{equation}
r_l = r_0 \exp \Bigl[-\sum_{k=1}^{l}\frac{2}{D_k-(2k-1)
S_k}e^{-\Phi(r_{k-1})}\Bigr] \label{rla}
\end{equation}
or combining with (\ref{phitsol}),
\begin{eqnarray}
r_l& = & r_0\exp\Bigl[- \Bigl(\sum_{k_1=1}^{l}\frac{2}{D_{k_1}-(2k_1-1)
S_{k_1}}\nonumber\\
&&\Bigl(1+\sum_{k_2=1}^{k_1-1}\frac{2S_{k_2}}{D_{k_2}-(2k_2-1)
S_{k_2}}\Bigl(1+\sum_{k_3=1}^{k_2-1}\frac{2S_{k_3}}{D_{k_3}-(2k_3-1)
S_{k_3}}\nonumber\\
&&\Bigl(1+\sum_{k_4=1}^{k_3-1}\frac{2S_{k_4}}{D_{k_4}-(2k_4-1)
S_{k_4}}\Bigl(\cdots \Bigl(1+\frac{2S_{1}}{D_{1}-S_1}\Bigr)\Bigr)\Bigr)\Bigr)\Bigr)\Bigr) e^{-\Phi(r_0)}\Bigr]. \label{rlb}
\end{eqnarray}
For instance,
\begin{equation}
r_1 = r_0 \exp \Bigl[-\frac{2}{D_1-S_1}e^{-\Phi(r_0)}\Bigr], \label{r1}
\end{equation}
\begin{equation}
r_2 = r_0 \exp \Bigl[-\Bigl(\frac{2}{D_1-S_1}+\frac{2}{D_2-3
S_2}\Bigl(1+\frac{2S_1}{D_1-S_1}\Bigr)\Bigr)e^{-\Phi(r_0)}\Bigr].\label{r2}
\end{equation}

Next we would like to find the equation of motion for the warp factor. We will do that without a need for computing the explicit expressions for the fluxes, since it is actually $H^{3/2} H_3^2$  which will come in the equation of motion of the warp factor  after combining (\ref{dphi}) and (\ref{traceR}). If one needs to determine the fluxes explicitly, then it would be necessary to decompose the fluxes with appropriate ansatz and solve all the supergravity equations of motion consistently.  First let us rewrite (\ref{dphi}) and (\ref{traceR}) with $C_0=0$, the metric given by (\ref{10dmetric}) and the corresponding Ricci scalar (\ref{ricci106}),
\begin{equation}
({\Phi''+\frac{5}{r}\Phi'}){\h}=
\frac{1}{12}\left(e^{\Phi}g_s^2\bar{f}-e^{-\Phi}\bar{h}\right),
\label{hgensola}
\end{equation}
and
\begin{equation}
-({H''+\frac{5}{r}H'})= \Phi'^2 H+
\frac{1}{12}\left(e^{\Phi}g_s^2\bar{f}+e^{-\Phi}\bar{h}\right),
\label{hgensolb}
\end{equation}
where we have defined
\be
\bar{f}\equiv {\h}^{3/2}F_3^2,\qquad \bar{h}\equiv {\h}^{3/2}H_3^2.\label{FHHdef}
\ee
Whenever the dilaton runs, as in the case here, both sides of (\ref{hgensola}) are nonzero and consequently $\star F_3 \ne e^{-\Phi}H_3/g_s$ and the 3-form combination $F_3-i e^{-\Phi}H_3/g_s$ is not imaginary self dual.
We have already found $\bar{h}$ from the gauge/gravity duality mapping, with (\ref{H3sol}) and (\ref{FHHdef}),
\begin{eqnarray}
\bar{h} = \frac{27 \pi^2\alpha'^2}{r^6} \Bigl(D_l+
S_l \bar{b}_2(r) \Bigr)^2 e^{2 \Phi}.\label{H3phib2a}
\end{eqnarray}
Now we can use (\ref{hgensola}) to find $e^{\Phi}g_s^2\bar{f}$ and substitute it into (\ref{hgensolb}) to find the equation of motion for the warp factor in terms of known quantities,
\be
H''+\frac{5}{r} H'+(\Phi'^2+\Phi''+\frac{5}{r}\Phi') H= -\frac{1}{6}e^{-\Phi}\bar{h}.\label{hgensolda}
\ee
The form of the equation of motion given by (\ref{hgensolda}) holds for any $H_3$ flux so far as all its components are in the internal (extra) space. Because $F_3$ drops out in combining (\ref{hgensola}) and (\ref{hgensolb}), it could have any components (and the $\bar{f}$ terms could have additional factors of $H$) so far as it is consistent with $H_3$ having internal components and the fluxes satisfy the Bianchi identity and the supergravity equations.
The equation of motion (\ref{hgensolda}) remains valid, even if we turn on $F_1$, since it drops out in the
same way as $F_3$ too.
Moreover, it is always possible to express the fluxes with all components in the internal space using the democratic formulation \cite{Bergshoeff:2001pv}.
We know from the running of the dilaton (\ref{t1p2phie}) that
\be
\Phi'=\frac{S_l}{r}e^{\Phi},\qquad \Phi''=\frac{S_l^2}{r^2}e^{2\Phi}-\frac{S_l}{r^2}e^{\Phi}\label{phippp}
\ee
and (\ref{hgensolda}) can be rewritten as
\be
H''+\frac{5}{r} H'+\frac{2}{r^2}(S_l^2e^{2\Phi}+2S_l e^{\Phi}) H=
-\frac{1}{6}e^{-\Phi}\bar{h}.\label{hgensoldf}
\ee
Thus we have obtained the equation of motion for the warp factor which contains only variables we have determined in the gauge/gravity duality mapping.

Let us write $\bar{h}$ and the warp factor $H$ as
\be
\bar{h}=\bar{h}_0+\delta{\bar{h}}, \qquad H=H_0+\delta{H},\label{dhdHdefn}
\ee
where $\bar{h}_0={243\alpha'^2 g_s^2M^2}/{r^6}$
and $H_0$ given by (\ref{warpH0}) are the corresponding values in the absence of corrections to the anomalous dimension. The equation of motion for the correction term is then
\be
\delta H''+\frac{5}{r} \delta H'=W_l,\label{Hdeqn}
\ee
where
\be
W_l\equiv-\frac{1}{6}(e^{-\Phi}\bar{h}-\bar{h}_0)-\frac{2}{r^2}(S_l^2e^{2\Phi}+2S_l e^{\Phi}) H_0\label{Adefn}
\ee
Integrating (\ref{Hdeqn}) twice,
\be
\delta H(r)=\delta H(r_0)+\int_{r_0}^{r}\frac{dr'}{r'^5}\Bigl( r_0^5\delta H'(r_0)+\int_{r_{0}}^{r'}dr''r''^5W_l(r'')\Bigr),\label{dHinteqn}
\ee
where the boundary values $\delta H(r_0)$ and $\delta H'(r_0)$ need to be fixed appropriately such that $H=H_0$ when $K\to \infty$ and $H|_{M=0}=27 \pi \alpha'^2 g_s N/{4 r^4}$ from the $N$ $D$3-flux.

Here we will calculate the warp factor perturbatively to leading $1/K$ corrections.  The expressions we present in the remaining part of this section contain only leading corrections in $1/K$ and higher order terms are ignored.
Expanding
the terms that come in (\ref{hgensoldf}) or (\ref{Adefn}) for the flow from the $(l-1)^{th}$ to the $l^{th}$ duality transition locations,
\be
S_l^2e^{2\Phi}+2 S_l e^{\Phi}=\frac{3g_s M}{\pi}\frac{1}{K},\label{term3a}
\ee
\ba
e^{-\Phi}\bar{h}=\frac{243\alpha'^2 g_s^2
M^2}{r^6}\Bigl[1+ \Bigl(\frac{9g_s M}{2\pi}\ln(\frac{r}{r_0})-\frac{3g_s M}{\pi}\ln(\frac{r_{l-1}}{r_0})\nonumber\\-(2l-3)\Bigr)\frac{1}{K}
\Bigr].\label{fir1a}
\ea
With (\ref{term3a}) and (\ref{fir1a}) in (\ref{Adefn}), we have
\be
W_l= -\frac{A \ln(\frac{r}{r_0})+B_l}{r^6}\frac{1}{K},\label{Wldefa}
\ee
where
\be
A=\frac{243\alpha'^2 g_s^3 M^3}{\pi},
\ee
\be
B_l=\frac{243\alpha'^2 g_s^2 M^2}{\pi}\left(-\frac{g_s M}{2}\ln(\frac{r_{l-1}}{r_0})+\frac{(K-(2l-3))\pi}{6}+\frac{g_s M}{16}\right)
\ee
or, equivalently, (\ref{term3a}) and (\ref{fir1a}) with (\ref{dhdHdefn}) in (\ref{hgensoldf}) give the equation of motion for leading order corrections,
\ba\delta H''+\frac{5}{r}\delta H'+ \frac{6g_s M}{\pi r^2}\frac{1}{K}H_0=-\frac{81\alpha'^2 g_s^2M^2}{2r^6}\Bigl(\frac{9g_s M}{2\pi}\ln(\frac{r}{r_0})\nonumber\\-\frac{3g_s M}{\pi}\ln(\frac{r_{l-1}}{r_0})-(2l-3)\Bigr)\frac{1}{K}.\label{hgensolbb}\ea
The warp factor containing leading $1/K$ corrections can then be obtained either by putting (\ref{Wldefa}) in (\ref{dHinteqn}) and integrating,  or by solving (\ref{hgensoldf}) with the boundary condition such that $H|_{M=0}=27 \pi \alpha'^2 g_s N/{4 r^4}$ from the $N$ regular $D$3-branes and  $H=H_0$ when $K\to \infty$. The result is
\ba {H}& = & \frac{27 \pi \alpha'^2}{4 r^4}\Bigr[
g_s N+\frac{3 g_s^2 M^2}{2\pi}  [\ln
   (\frac{r}{r_0})+\frac{1}{4}]\nonumber\\&&-
   \Bigl(\frac{3 g_s^2 M^2}{2\pi}[\ln(\frac{r}{r_{l_s}})+\frac{1}{4}][\frac{3g_sM}{\pi}
   \ln(\frac{r_{l-1}}{r_0})+(2l-3)-\frac{3g_s M}{8 \pi}]\nonumber\\&&-\frac{9 g_s^3 M^3}{4\pi^2}  [\ln
   (\frac{r}{r_0})+{2}(\ln(\frac{r}{r_0}))^2+\frac{1}{4}]
   \Bigr)\frac{1}{K}\Bigr]\label{HdirDE}
\ea
for the flow from the $(l-1)^{th}$ to the $l^{th}$ duality transition locations with appropriate choice of boundary conditions and  we will take $r_{l_s}$ to be the value of $r$ where the size of a step is being estimated. Note that a step in the warp factor comes from terms in the second line in (\ref{HdirDE}) which is linear in $\ln r$.

\begin{figure}[t]
\begin{center}
\leavevmode
\includegraphics[width=0.7\textwidth, angle=0]{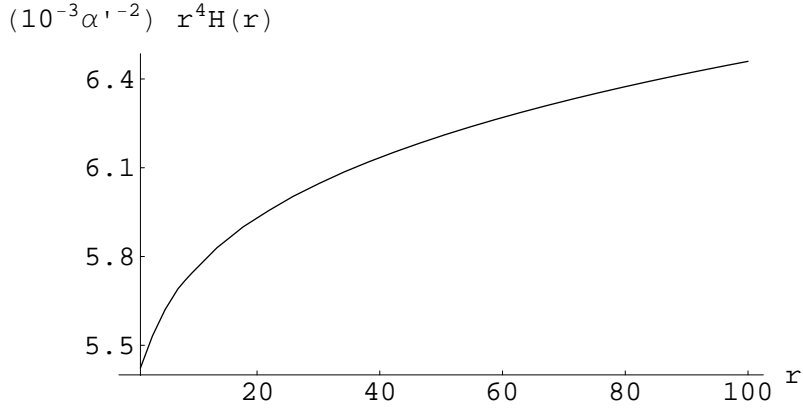}
\caption{Warp factor plotted for parameters  $g_s=0.3$, $K=100$, $M=20$, $c=0.0001$ and $r_0=100$.}
\label{Hplot}
\end{center}
\end{figure}

\begin{figure}[t]
\begin{center}
\leavevmode
\includegraphics[width=0.65\textwidth, angle=0]{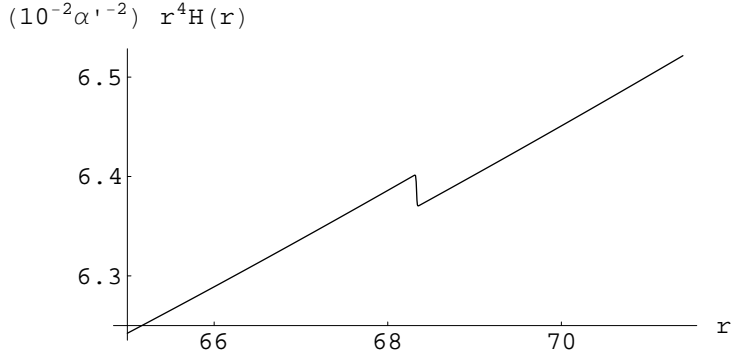}
\caption{A magnified plot of the first step at $r=r_1$ in the warp factor for parameters  $g_s=0.3$, $K=5$, $M=20$, $c=0.0001$ and $r_0=100$.}
\label{Hstep1}
\end{center}
\end{figure}

To get an estimate of the sizes of the steps, let us consider the step at $r=r_1$, where $r_1$ is given by (\ref{r1}), to leading order in $1/K$,
\be
\ln (\frac{r_{1}}{r_{0}}) \simeq - \frac{2 \pi}{3 g_{s}M} \left(1+ \frac{1}{2K} \right)
\ee
The step in the warp factor at $r_1$ comes from the difference between the warp factor $H(1,r)$ for the flow from $r_0$ to $r_1$ and the warp factor $H(2,r)$ for the flow from $r_1$ to $r_2$. This follows from (\ref{HdirDE}),
\ba {\h}(1,r_1)-\h(2,r_1)& \simeq &
   \frac{27 \pi \alpha'^2}{4 r_1^4}\Bigr(\frac{3 g_s^2 M^2}{2\pi}[\frac{1}{4}][\frac{3g_sM}{\pi}
   \ln(\frac{r_{1}}{r_{0}})+2]\Bigr)\frac{1}{K}\nonumber\\&\simeq&
-\frac{27\pi \alpha'^2g_s N}{4 r_1^4}\left(\frac{3g_sM}{8\pi}\right)\frac{1}{K^3}.\label{Hdiffstep1}\,\,\,\,\,\,\,
\ea
and the relative size of the step is
\ba \frac{{\h}(1,r_1)-\h(2,r_1)}{{\h}(1,r_1)}\simeq -\frac{3g_sM}{8\pi}\frac{1}{K^3}.\label{Hdiffstep1b}\,\,\,\,\,\,\,
\ea
Thus the height of the step in the warp factor is of order $1/K^3$. The reasons for the $1/K^3$ order in the height of the step are the following. First, the correction in $H(r)$ itself comes at order $1/K$. Second, the step comes in the difference between $H(1,r)$ and $H(2,r)$ across a duality transition which gives an additional factor $\propto g_sM/K$ effectively coming from corrections to the location of the step. These two together give a factor $\propto (1/K)(g_sM/K)=g_s N/K^3$. We also note that
the warp factor
steps up
as the theory flows across $r=r_1$ down toward the bottom of the throat.
We connect the steps with interpolating $tanh$ function and
write the warp factor given by (\ref{HdirDE}) as
\begin{equation}
H(r) =  H(1,r)+\sum _{l=1}^{K-1}\frac{1}{2} \Bigl(H(l+1,r)- H(l,r) \Bigr)
\Bigl(1+\tanh \Big[ \frac{r_{l}-r}{ r_{l}c} \Big]\Bigr),\label{Hrdef2}
\end{equation}
where $ H(l,r)$ is the warp factor given in (\ref{HdirDE}) for the flow in the region from the $(l-1)^{th}$ to the $l^{th}$ duality transition locations and $c$ is a parameter which describes the sharpness of the steps.
An illustration of the warp factor is shown in Figures \ref{Hplot} and \ref{Hstep1}. The steps are not visible in Figure \ref{Hplot}, since we have taken large $K$. A magnified plot of the first step in the warp factor is shown for a relatively small value of $K$ in Figure \ref{Hstep1}.

Now we want to discuss how we can isolate and focus on some number of steps and write down the warp factor for a flow including only $s$ number of steps starting from some location of duality transition at $r=r_{l_0}$; i.e, in the range $r_{l_0-1}<r<r_{l_0+s}$.
Since we may be in a region far away from the edge of the throat, we use the expansion parameter $1/K_{\mathrm{eff}}=M/N_{\mathrm{eff}}=1/(K-(l-1))$, where $N_{\mathrm{eff}}=N-(l-1)M$ is the effective number of D3-brane charge for the flow between $r_{l-1}$ and $r_l$.
The warp factor (\ref{Hrdef2}) in this case can then be written as
\begin{equation}
H(r) =  H(l_0,r)+\sum _{l=l_0}^{l_0+s-1}\frac{1}{2} \Bigl(H(l+1,r)- H(l,r) \Bigr)
\Bigl(1+\tanh \Big[ \frac{r_{l}-r}{c r_{l}} \Big]\Bigr),\label{Hrdef2l}
\end{equation}
where $H(l,r)$ is given by (\ref{HdirDE}) and we set $l=l_0$ for $H(l_0,r)$.
A step in the warp factor at $r_l$  comes from the difference $H(l+1,r_l)- H(l,r_l)$ and we have
from (\ref{HdirDE}),
\ba {\h}(l,r_l)-\h(l+1,r_{l+1})& \simeq &
   \frac{27 \pi \alpha'^2}{4 r_l^4}\Bigr(\frac{3 g_s^2 M^2}{2\pi}[\frac{1}{4}][\frac{3g_sM}{\pi}
   \ln(\frac{r_{l+1}}{r_{l}})+2]\Bigr)\nonumber\\&&\times \frac{1}{K-(l-1)}.\label{Hdiffstepl}
\ea
We also have from (\ref{rrel})
\be
\log(\frac{r_l}{r_{l-1}})\simeq -\frac{2\pi}{3g_sM}\left(1+\frac{1}{2(K-(l-1))}\right)\label{logrlrlm1}
\ee
with the string coupling normalized such that $e^{\Phi(r_{l-1})}$ is absorbed in $g_s$ here.
The step in $H(r)$ at $r_l$ is then obtained using (\ref{logrlrlm1}) in (\ref{Hdiffstepl}),
\begin{eqnarray} H(l,r_l)-H(l+1,r_l)& \simeq & -
\frac{27\pi \alpha'^2 g_s (K-(l-1))M}{4 r_l^4}\left(\frac{3g_sM}{8\pi}\right)\nonumber\\&  &\times \frac{1}{(K-(l-1))^3},\label{Hdiffsteplb}
\end{eqnarray}
where $(K-(l-1))M=  N_{\mathrm{eff}}$, is the effective $D3$-branes charge for the flow from the $(l-1)^{th}$ to the $l^{th}$ duality transition point.
The warp factor which describes the region with $s$ number of steps after and including the step at $r=r_{l_0}$ is then
\begin{eqnarray}
H(r) & \simeq & H(l_0,r)
  + \frac{27\pi \alpha'^2}{4 r^4}\sum_{l=l_0}^{l_0+s-1}\Bigl[g_s (K-(l-1)) M\left(\frac{3g_sM}{8\pi}-1)\right)\nonumber\\&  &\times \frac{1}{2(K-(l-1))^3}
\Bigl(1+\tanh [ \frac{r_{l}-r}{c r_{l}} ]\Bigr)\Bigr].\label{Hrdef2b}
\end{eqnarray}

\section{$D$3-brane world volume effective action}

In this section we like to discuss the $D$3-brane world volume effective action in the warped geometry. In particular, we like to see how the steps in the warp factor from the Seiberg duality cascade show up here.
The presence of steps is actually very generic as a correction to the approximate
geometry for a warped throat. The action here may be used to study
how brane inflation can be implemented such that the steps may give possible stringy signatures in the cosmic microwave background radiation in the KKLMMT inflationary scenario. To be concrete, we shall present our discussion within the context of the KS throat discussed above.


In the KKLMMT scenario, inflation takes place when a $D$3-brane moves in the throat. The inflaton is the location of the $D$3-brane from the bottom of the throat. The $D3$-potential gets contributions from a Dirac-Born-Infeld (DBI) term and a Chern-Simons term in the action, as well as possible terms due to a presence of $\D$3-brane sitting at the bottom of the warped deformed throat. The DBI term contains $e^{-\Phi}$ while the Chern-Simons term does not. When the dilaton is constant, as in the KS solution, the two terms cancel out and the $D3$-potential vanishes in the absence of a $\D$3-brane. Consequently, one needs other sources such as $\overline{D}$3-brane to attract the $D$3-brane. Here we can have a dilaton driven inflation, since the dilaton runs and the $D$3-potential is dynamically nonzero.

Presumably, we are interested in inflation that takes place when $r
< r_{0}$, where $r_0$ is the location of the edge of the throat.
Thus, given $r_{0}$, we can determine all the other duality
transition locations $r_l$ and the bottom of the throat is at $r_A \approx
r_{K}$. In the large $K$ limit and for small
$l$, the corrections may be very small.

Including the expansion of the universe, the 10-dimensional metric takes the form:
\be
ds^2={\h}^{-1/2}(r)(-dt^{2}+a(t)^{2} d{\bf x}^{2})
+ {\h}^{1/2}(r)(dr^2+r^2 ds_{T^{1,1}}^2)
\label{10dmeta}
\ee
Here the cosmic scale factor $a(t)$ is that of an expanding homogeneous isotropic universe spanned by the 3-dimensions ${\bf x}$,
and $r$ is the coordinate for the flow in the throat. The metric $ds_{T^{1,1}}^2$ is for
base of the conifold. Warped spaces are natural in string theory models and are useful for flattening potentials and for generating a hierarchy of scales with the UV at the top (edge) of the throat and the IR scale at the warped bottom (around $r \sim r_A$). Crudely, ${\h}(r)\sim (r/R)^{-4}$, where $R \gg r_A$ is the scale of the throat.
The expression for the action includes the dilaton $\Phi$, the metric
$G_{MN}$, the anti-symmetric tensor $B_{MN}$ and the gauge field $F_{MN}$. The Chern-Simons term contains couplings between the brane and R-R fields ($p$-forms) $C_p$, with $p$ even for type IIB theory. We use variable $\xi$ and indices $\{a,b\}$ for coordinates and quantities on the brane and we have
\ba
\label{fullaction}
S_{D3}&=&-T_3\int d^4\xi e^{-\Phi}\:\sqrt{det |G_{ab}+B_{ab}+2\pi\ap F_{ab}|}\\\nonumber
&&\pm \mu_3\int_{\mathcal{M}_4}\left[\sum_{p=0}^4C_{p}\right]\wedge \mathrm{tr} \left[e^{2\pi\ap F+B}\right]
\ea
where $T_3=[(2\pi)^3g_s\alpha^{\prime 2}]^{-1}$ is the tension of a $D$3-brane. Here $g_{s}$ is the string coupling and the Regge slope $\alpha^{\prime}=m_{s}^{-2}$ sets the string scale, where $m_{s}$ is the string mass scale.
The quantities inside the determinant have been pulled back onto the brane and the $\pm$ is for a brane/anti-brane. In particular, note that the pulled back metric ($G_{ab}$) will contain the warp factor. In the second term $\mu_3=g_sT_3$ is the brane charge. The integration is over the $D$3-brane world volume, and contributions should, of course, have the correct dimension. In the simplest cases, many of the fields appearing in (\ref{fullaction}) are trivially zero. However, for more interesting solutions we will need to solve the supergravity equations to find each of the fields.

We now consider the simple case with vanishing pullbacks $B_{ab}$ and $F_{ab}$, and only $C_4$ non-zero ($D$3-branes are charged under the 4-form RR field $C_{4}$). The $D$3-brane alone in this background is supersymmetric. In addition, the supergravity equations require that components of $C_4$ on the brane be $1/{\h}(r)g_s$. We have aligned the brane coordinates with the usual space-time coordinates so that the only non-zero derivatives in the pullback are those with respect to time, and we assume that only the radial motion of the brane is important. Then (\ref{fullaction}) simplifies to
\be
S_{D3}=\int d^4x  \frac{a^3(t)T_3}{{\h}(r)} \left(- e^{-\Phi} \sqrt{1- {\h} (r) \dot{r}^2} + 1 \right).
\ee
For slow motion along $r$, this reduces to
\be
S_{D3} \approx  \int d^4x  a^3(t)T_3 \left( e^{-\Phi} \frac{\dot{r}^2}{2} - \frac{1}{{\h}(r)} ( e^{-\Phi} -1) \right).
\label{srr}
\ee
We see that the $D$3 potential $V_{D3}(r)= (e^{-\Phi(r)}-1)/{{\h}(r)}$ vanishes for constant dilaton $\Phi=0$.

The inflaton $\phi$ is related to the position of a space-time filling $D$3-brane moving in such a throat. Specifically,
\be
\phi=\sqrt{T_3}r
\ee
so we have the following inflaton action in the slow-roll regime,
\begin{equation}
S_{D3} \approx \int dx^{4} a^{3}(t)  \left[e^{-\Phi (\phi)}\frac{{\dot \phi}^{2}}{2} - \frac{T_{3}}{{\h}(\phi)}(e^{-\Phi (\phi)} -1) \right],\label{SD3a}
\end{equation}
where $\Phi (r) \rightarrow \Phi (\phi)$ and ${\h} (r) \rightarrow {\h} (\phi)$.

The kinetic term in this action has the form considered in \cite{Shiu:2001sy}.
The potential is similar to that considered in \cite{Dymarsky:2005xt}. Effects of features in the inflaton potential were also studied in \cite{Ashoorioon:2006wc}.
Note that the $D$3-potential term in (\ref{SD3a}) vanishes for $\Phi=0$. One may consider two possibilities:
(1)  $\Phi=0$ as the asymptotic value of the dilaton at large $r$. That is, the $D$3-brane is mobile in the bulk away from the throat;
(2)  $\Phi=0$ at the bottom of the throat, so the $D$3-brane is BPS there.

Here we take $\Phi=0$ at $r=r_0$ or $1+S_1c_1=0$.
We can always add an inflaton mass term if necessary.
Using (\ref{phitsol}),
\begin{equation}
\frac{d}{d t} \,e^{-\Phi} =-S_{l} \frac{\dot r}{r} = -S_{l} \frac{\dot \phi}{\phi}.
\end{equation}
Now we connect the steps in the $D$3-brane potential with an interpolating $tanh$ function. First
let us write the $D$3-brane potential for the flow between $r_{l-1}$ to $r_l$ as
\be
V_{D3}(l,r)= T_3\frac{e^{-\Phi(l,r)}-1}{{\h}(l,r)},\label{vd31}
\ee
where we use $\h (l,r)$ given by (\ref{HdirDE}),
\ba
{H(l,r)}& = & \frac{27 \pi \alpha'^2}{4 r^4}\Bigr[
g_s N+\frac{3 g_s^2 M^2}{2\pi}  [\ln
   (\frac{r}{r_0})+\frac{1}{4}]\nonumber\\&&-
   \Bigl(\frac{3 g_s^2 M^2}{2\pi}[\ln(\frac{r}{r_{l_s}})+\frac{1}{4}][\frac{3g_sM}{\pi}
   \ln(\frac{r_{l-1}}{r_0})+(2l-3)-\frac{3g_s M}{8 \pi}]\nonumber\\&&-\frac{9 g_s^3 M^3}{4\pi^2}  [\ln
   (\frac{r}{r_0})+{2}(\ln(\frac{r}{r_0}))^2+\frac{1}{4}]
   \Bigr)\frac{1}{K}\Bigr],\label{hsoln-l}
\ea
and $\Phi$ given by (\ref{phitsol1}, \ref{phitsol}),
\be
e^{-\Phi(l,r)}=1+S_{1}c_{1} -\sum_{k=1}^{l-1}S_{k}\ln(r_k/r_{k-1})
-S_{l}\ln(r/r_{l-1}). \label{phitsol-l}
\ee
We then connect the steps with interpolating $tanh$ function and
write the $D3$-brane potential given by (\ref{vd31}) as
\begin{equation}
V_{D3}(r) \approx  V_{D3}(1,r)+\sum _{l=1}^{K-1}\frac{1}{2} \Bigl(V_{D3}(l+1,r)- V_{D3}(l,r) \Bigr)
\Bigl(1+\tanh \Big[ \frac{r_{l}-r}{ r_{l}d} \Big]\Bigr),\label{Frdef2}
\end{equation}
where again $K\equiv N/M$.  We expect the steps to be smoothed out by the scale at the Seiberg duality transition, namely, $\Lambda_{l} \sim r_{l}$, so we introduce the width $d_{l}=r_{l}d$, where the parameter $d$ measures the sharpness of the steps, with smaller $d$ corresponding to sharper steps. As the theory flows to smaller $AdS_5$ radius toward the bottom of the
throat, $r_{l}$ decreases, so the steps are becoming sharper.
\begin{figure}[t]
\begin{center}
\leavevmode
\includegraphics[width=0.75\textwidth, angle=0]{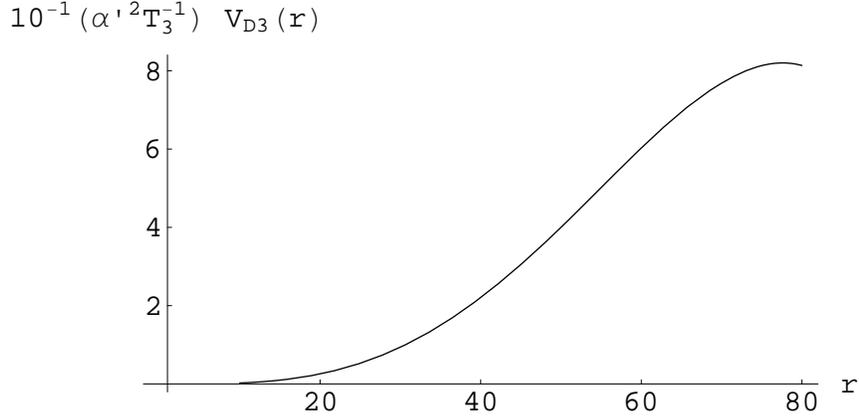}
\caption{$D3$-brane potential plotted for parameters  $g_s=0.3$, $K=100$, $M=20$, $d=0.0001$ and $r_0=100$.}
\label{potentialsteps}
\end{center}
\end{figure}

The height of the first step in the potential follows from (\ref{Hdiffstep1}) and (\ref{vd31}),
\ba
V_{D3}(1,r_1)-V_{D3}(2,r_1)&=& T_3        \Bigl(\frac{H(2,r_1)-H(1,r_1)}{H(1,r_1)H(2,r_1)}\Bigr)(e^{-\Phi(r_1)}-1)\nonumber
\\&\simeq&
\frac{4T_3r_1^4}{27\pi \alpha'^2 g_sN} \left(\frac{3g_sM}{8\pi}\right) \frac{1}{K^4}
\ea
and the absolute magnitude of the step is of leading order $1/K^4$.
The relative change in the potential at $r=r_1$ is
\ba
\frac{ \delta V_{D3}}{V_{D3}} = \Bigl(\frac{V_{D3}(1,r_1)-V_{D3}(2,r_1)}{V_{D3}(1,r_1)}\Bigr)&=&         \Bigl(\frac{H(2,r_1)-H(1,r_1)}{H(2,r_1)}\Bigr)\nonumber\\&\simeq &
\left(\frac{3g_sM}{8\pi}\right)\frac{1}{K^3}.
\ea
and the relative size of the step is of order $1/K^{3}$. 
Similarly, the height of the $l^{th}$ step in the potential is
\ba
V_{D3}(l,r_{l})-V_{D3}(l+1,r_{l})&=& T_3        \Bigl(\frac{H(l+1,r_l)-H(l,r_l)}{H(l,r_l)H(l+1,r_l)}\Bigr)(e^{-\Phi(r_l)}-1)\nonumber
\\&\simeq&
\frac{4T_3r_1^4}{27\pi \alpha'^2 g_s(K-(l-1))M} \left(\frac{3g_sM}{8\pi}\right) \nonumber
\\&&\times\frac{1}{(K-(l-1))^4}
\ea
An illustration of the $D$3-brane potential $V_{D3}$ is shown in Figures \ref{potentialsteps}. The steps are not visible on the plot, since we have taken large $K$. The potential we have here
steps down
as the theory flows across $r=r_l$ down toward the bottom of the throat.

Since the step sizes are $\mathcal{O}(1/K^{4})$ while the dilaton running starts at $\mathcal{O}(1/K)$, it is useful to consider the potential without the steps. For $r_0 \ge r \ge r_{1}$,
\ba
V_{D3}(\phi) &\simeq &   \frac{ - 32 \pi^{2}\phi^{4}\left( \ln (\phi/\phi_{0}) - 1/4 -1/16\bet \right)}{27 KM [ \bet  + \ln (\phi/\phi_{0}) +1/4]},
\ea
where
$$\bet =2\pi K/3 g_{s}M.$$
Note that we have fixed the constant in (\ref{phitsol1}) to be $c_{1}=1/4 + 1/16\bet$ so that $V_{D3}(\phi)$ is monotonic in the range we are interested in.
$$\phi_{A} \simeq \phi_{K }= \phi_{0}e^{-2\pi K/3 g_{s}M}= \phi_{0}e^{-\bet}$$
$$H(\phi_{K})=e^{4 \bet}/4 \bet $$
so at the edge,
$$V_{D3} (\phi_{0})=  T_{3}/4 \bet.$$

There are 5 parameters here: $\alpha^{\prime}$, $g_s$, $K$, $M$ and $d$.
It is also reasonable to expect that the $D$3-brane is BPS at the bottom of the throat, $\phi=\phi_A$. In this case, $\Phi_A=\Phi (\phi_A)=0$ and $c$ would be determined in terms of the other parameters. Here we have set $\Phi=0$ at the edge of the throat, $r=r_0$.

If the inflaton is moving relativistically, the NS-NS term in $V_{D3} (\phi)$ becomes part of the DBI kinetic action and so $V_{D3} (\phi) \rightarrow -T_{3}/H$. This interesting case should be studied carefully.

As the $D3$-brane moves across a step, it would generate oscillations in the angular power spectrum of density perturbations. Moreover, for $K$ big enough such that we could see two or three steps, it might be possible to correlate the steps and make some definitive predictions.
In comparing with \cite{Dymarsky:2005xt}, we see that the warp factor and so $V_{D3}(\phi)$ can be quite sensitive to the details of the throat geometry. This fact should be taken into account in any model comparison to cosmological data.

In the $D$3-$\D$3-brane inflationary scenario, there is a $\D$3-brane sitting at the bottom of the throat at $\phi=\phi_A$. So we should introduce $V_{D{\bar D}}(r)$; this is independent of $\Phi$, so it is not modified from the KS case.
\ba
V(\phi) &=& V_{D \bar D}(\phi) + V_{D3}(\phi), \\\nonumber
V_{D \bar D}(\phi) &=& V_{0}\left(1-\frac{27}{64\pi^{2}}\frac{V_{0}}{\phi^4}\right), \\ \nonumber
V_{0} &=& 2 T_{3}/H(\phi_{K})=8 T_{3} \bet e^{-4 \bet},
\ea
where the constant term $V_{0}$ is the effective vacuum energy. Additional terms coming from the K\"{a}hler potential may be present. Such terms may also have strong dependence on the dilaton. Back-reaction may ameliorate the size and/or the steepness of the potential. This requires a more careful analysis.

\section{The warped deformed/resolved conifold}

The deformation of the conifold via gaugino condensation in the
gauge theory in the KS throat is related to a geometric transition in the gravity
theory where the $S^2\subset T^{1,1}$ cycle shrinks to zero size,
the $M$ number of $D5$ branes wrapping $S^2$ disappear and are
replaced by flux through $S^3\subset T^{1,1}$. Thus the tip of the
deformed conifold is $S^3$. The metric which describes the deformed
conifold thus involves an interpolation between $T^{1,1}$ at large
$r$ and $S^3$ at the tip of the conifold. However, it was later discussed that
a flow to a baryonic branch with a quantum deformed moduli space of $SU(2M)\times SU(M)$, in the case where $N$ is an integral multiple of $M$, might be the preferred route of the flow \cite{Gubser:2004qj, Butti:2004pk}.
In the MN case, $\mathcal{N}=1$ supersymmetric gauge theory was obtained by wrapping NS5-branes on $S^2$.
A metric and flux ansatz which could give an interpolating solution between KS and MN was put forward in \cite{Papadopoulos:2000gj}. A leading order perturbative expansion around the KS solution was found in \cite{Gubser:2004qj}. Later,  $SU(3)$ structures were used to find a one parameter set of solutions which flow in a direction from KS to MN \cite{Butti:2004pk}.
We used Einstein frame
in previous sections. In this and the next section, we will be
using the string frame and also absorb $g_s$ in $e^{\Phi}$. The metrics in the two
frames are related by
$G_{MN}(\mathrm{string})=e^{\Phi/2}G_{MN}(\mathrm{Einstein})$.

\subsection{SU(3) structures}

In this section we will briefly review the basic ideas in applying $SU(3)$
structures to study supergravity backgrounds with torsions. The study of supersymmetry conditions for supergravity backgrounds with torsion was initiated by Strominger \cite{Strominger:1986uh}. See \cite{Gauntlett:2002sc, Gurrieri:2002wz,Gauntlett:2003cy,Butti:2004pk} for details on applying group structures to supergravity.

Consider a compactification of type IIB strings on $R^{(1,3)}\times Y$, where $Y$ is a compact six dimensional manifold. The Clifford algebra in ten dimensions is described by ten $32\times 32$ gamma matrices. Let us denote these gamma matrices by $\Gamma^M$, where the uppercase letters $M,N,\cdots$ run over $0,1,\cdots, 9$. The gamma matrices satisfy $\{\Gamma^{M}, \Gamma^{N}\}=2 G ^{MN}$,
where $G _{MN}$ is the metric. The generators of the Lorentz group $Spin(1,9)$ on $R^{(1,3)}\times Y$ can be constructed
as commutators of the gamma matrices.
The spinor representation in 10-d is given by $\Gamma_{(10)}=\Gamma^0 \Gamma^1\cdots \Gamma^9$.
The Lorentz algebra decomposes to $Spin(1,3)\times Spin(6)$ on $R^{(1,3)}\times Y$.
We can write $\Gamma_{(10)}=\Gamma_{(4)}\Gamma_{(6)}$, where
$\Gamma_{(4)}=-i \Gamma^{0}\cdots\Gamma^{3}$ and $\Gamma_{(6)}=i
\Gamma^{4}\cdots\Gamma^{9}$ denote the spinor representations on $R^{(1,3)}$ and on $Y$ respectively.
There are two spinors of the same chirality in IIB which decompose under $Spin(1,3)\times Spin(6)$ as
$\epsilon^1=\zeta_{+}\eta^{1}_{+}+\zeta_{-}\eta^{1}_{-}$ and $\epsilon^2=\zeta_{+}\eta^{2}_{+}+\zeta_{-}\eta^{2}_{-}$, where $\zeta_{+}$ is the spinor on $R^{(1,3)}$, $\eta^{i}$ are the spinors on $Y$, $\zeta_{-}={\zeta_{+}}^{*}$, and $\eta^{i}_{-}={\eta^{i}_{+}}^{*}$.
The number of supersymmetries in 4-d depends on the structure group on $Y$. A generic $Y$ with structure group
$SO(6)\sim SU(4)$ has no globally defined covariantly constant spinor and gives no supersymmetry. The spinor representation of $SO(6)$
corresponds to the fundamental
representation of $SU(4)$ which decomposes as $1\oplus3$ under $SU(3)$.
Thus there is one globally defined $SU(3)$ singlet spinor on $Y$.
In order to preserve some supersymmetry, $Y$ needs to have a reduced structure group and to preserve $\mathcal{N}=1$ supersymmetry the structure group on $Y$ has to be reduced at least to $SU(3)$. In that case, if we denote the one $SU(3)$ singlet spinor mentioned above by $\eta_+$, the two spinors $\eta_+^{1,2}$ are complex proportional and
are related to the invariant
spinor in terms of two
complex functions $\alpha$ and $\beta$ which can be expressed as
$\eta_+^1=\frac{1}{2}(\alpha+\beta) \eta_+$ and
$\eta_+^2=\frac{1}{2i}(\alpha-\beta) \eta_+$. If $Y$ is a Calabi-Yau threefold, the globally invariant spinor would also be covariantly constant
and depend trivially on the tangent frame bundle on $Y$ and these two spinors give $\mathcal{N}=2$ supersymmetry in four dimensions.

However, when fluxes are turned on, the geometry backreacts and
develops torsion and $Y$ could in general become
non-Ricci-flat and non-K\"{a}hler. When the extra space is compactified on a generalized
Calabi-Yau  with $SU(3)$ structures, the fluxes from the $N$
regular and $M$ fractional D3-branes give rise to torsions which
fall in various representations of $SU(3)$. In the presence of
fluxes, the spinor $\eta^+$ is not covariantly constant with respect
to the Levi-Civita connection but would be so with a connection
which includes torsion. The components of the torsion fall into the
$SU(3)$ representations
$(3+\bar{3})\otimes(3+\bar{3}+1)=(8+8)\oplus (6+\bar{6})\oplus
(3+\bar{3})\oplus (3+\bar{3})\oplus(1+1)$.
On the other hand, there are two $SU(3)$ singlets on $Y$, one is a fundamental 2-form which
describes the almost complex structure and the other is a globally
non-vanishing holomorphic 3-form. Unlike the case of Calabi-Yau
threefolds, these 2- and 3-forms are not closed now and the
different components of the torsion come in $dJ$ and $d\Omega$. $dJ$ has
$20$ components and decomposes under
$SU(3)$ as $(6+\bar{6})\oplus (3+\bar{3})+(1+1)$, and
$d\Omega$ transforms as a $24$ of $SU(4)$ and decomposes under
$SU(3)$ as $(8+8)\oplus (3+\bar{3})+(1+1)$. Similarly the fluxes can
be decomposed into different components in representations
of $SU(3)$. The different components of the torsion which fall in
representations of $SU(3)$ need to vanish or get balanced by fluxes
of the corresponding forms and representations in order to preserve
$\mathcal{N}=1$ supersymmetry. This gives constraints on the
relations among the parameters $\alpha$ and $\beta$, the fluxes and
the metric.

Next we want to see the torsion components in the variations of the fundamental 2-form and the holomorphic 3-form when $Y$ has $SU(3)$ structures. Suppose we have parameterized the metric on $Y$ as
\be\label{ds6s-1}
ds_6^2=\sum_{m=1}^{6}G_m^2,
\ee
where $G_m$ are real differential 1-forms which are not closed here. Lower case indices $m,n,\cdots$ run over $1$ to $6$ here. Let us then define
\be\label{ZiGi}
Z_1=G_1+i G_2,\quad Z_2=G_3+i G_4,\quad Z_3=G_5+i G_6.
\ee
We can then write the fundamental 2-form $J$ and the holomorphic 3-form $\Omega$ as
\ba\label{JZ-1}
J&=&
\frac{i}{2}  \sum_{i=\bar{i}=1}^{3}{Z_i}{ {\wedge} }{\bar{Z}_{\bar{i}}},
\ea
\ba\label{OmZ-1}
\Omega &=&
{Z_1}{ {\wedge} }{Z_2}{ {\wedge} }{Z_3}.
\ea
The $i$ is for holomorphic indices which run over $1$ to $3$, and the $\bar{i}$ is for the corresponding anti-holomorphic indices. However, the $Z_i$'s are not differentials of complex coordinates and we will need to impose constraints in order to make $Y$ a complex manifold.
Note that $J$ transforms as $(1,1)$ and $\Omega$ transforms as $(3,0)$. The complex and K\"{a}hler structures on $Y$ are determined by the properties in the variations of $J$ and $\Omega$. But it is easy to see that $dJ$ has components with forms $(2,1)\oplus (1,2)\oplus(3,0)\oplus(0,3)$ in the $Z_i$'s. Moreover, $d\Omega$ has components with $(3,1)\oplus (2,2)$ forms; it does not have a $(4,0)$, since a complex 4-form vanishes in three complex dimensions.  When $Y$ has $SU(3)$ structures, the components can further be broken down to representations of $SU(3)$. The $(3,0)\oplus(0,3)$ forms in $dJ$ fall in the singlet representation, the $(1,2)\oplus (2,1)$ forms fall in the $(6\oplus3)\oplus(\bar{6}\oplus\bar{3})$ representations. The $(3,1)$ form in $d\Omega$ falls in the $\bar{5}$ representation and the $(2,2)$ form falls in the $8\oplus1$ representations. All in all,
\be\label{dJ-1a}
dJ=-\frac{3}{2}\mathrm{Im}(W_1^{(1)} \bar{\Omega})+(W_4^{(3)}+W_4^{(\bar{3})})\wedge J+(W_3^{(6)}+W_3^{(\bar{6})}),
\ee
\be\label{dOm-1a}
d\Omega=W_1^{(1)} J^2+W_2^{(8)}\wedge J+W_5^{(\bar{3})}\wedge \Omega,
\ee
where the $W$'s denote components of the torsion.
If $Y$ is a Calabi-Yau manifold, then both $J$ and $\Omega$ are closed, $dJ=0$ and $d\Omega=0$, and all torsion components vanish. Thus nonvanishing components of the torsion measure the departure of the manifold from being Calabi-Yau.
The fluxes can also be decomposed as
\be\label{H3-1a}
H_3=-\frac{3}{2}\mathrm{Im}(H_3^{(1)} \bar{\Omega})+(H_3^{(3)}+H_3^{(\bar{3})})\wedge J+(H_3^{(6)}+H_3^{(\bar{6})}),
\ee
\be\label{F3-1a}
F_3=-\frac{3}{2}\mathrm{Im}(F_3^{(1)} \bar{\Omega})+(F_3^{(3)}+F_3^{(\bar{3})})\wedge J+(F_3^{(6)}+H_3^{(\bar{6})}).
\ee
If $Y$ is to be a complex manifold, the $(3,0)$ and $(0,3)$ components of $dJ$ and the $(2,2)$ components in $d\Omega$ must vanish which amount to demanding $W_1^{(1)}=0$ and $W_2^{(8)}=0$.

\subsection{Equations of motion in the $6\oplus \bar{6}$ sector}

The constraint on the relation between the fluxes and the torsions were found in  \cite{Butti:2004pk}. It will be enough for our purpose here to focus only on the equations of motion in the $6\oplus\bar{6}$ sector. In particular the equations of motion for flux and torsion components in the $6\oplus\bar{6}$ representations are the following three complex equations, of which only two are independent,
\be\label{W3F6}
(\alpha^2-\beta^2)W_3^{(6)}=2\alpha \beta e^{\Phi} F_3^{(6)},
\ee
\be\label{W3H6}
(\alpha^2+\beta^2)W_3^{(6)}=-2\alpha \beta  \star_6 H_3^{(6)},
\ee
\be\label{H6F6}
(\alpha^2-\beta^2)H_3^{(6)}=(\alpha^2+\beta^2) e^{\Phi} \star_6 F_3^{(6)}.
\ee

A one parameter of numerical solution for the supersymmetry conditions using the ansatz in \cite{Papadopoulos:2000gj} was obtained in \cite{Butti:2004pk} for the case of $\alpha$ real and $\beta$ imaginary, where the varying parameter arises from different possible values of the boundary value of dilaton at the very edge (or the very bottom) of the throat and the vacuum expectation value of the axionic scalar moduli field on the quantum deformed moduli space in the baryonic branch. We will see later that the supergravity side containing the corrections to the anomalous mass dimension from the gauge theory side does not fall into this solution. We will also see in the next section the implications of the corrections  in terms of $SU(3)$ structures.
For now, a simple way to see that the supergravity flow we have here is different is simply to note
that the leading order correction to the running of the dilaton in (\ref{phitsol}), if we just  consider the flow in the range $r_1\le r \le r_0$ and define $\tilde{t}\equiv \ln (r/r_0)$, comes at $\mathcal{O}(\tilde{t})$, which is different from the flow found in \cite{Gubser:2004qj} and \cite{Butti:2004pk}, where the leading order correction to the running of the dilaton comes at $\mathcal{O}(\tilde{t}^2)$. It will be important to construct the full dual supergravity background and flow corresponding to the supersymmetric gauge theory containing corrections to the anomalous mass dimension.

\section{Gravitational source for Seiberg duality transformations}

Now we want to see that the locations where Seiberg duality transformations occur have a geometric obstruction with a jump in the relation between the two complex proportional spinors $\eta_{+}^{1,2}$ on the six dimensional manifold $Y$. Conversely, this geometric obstruction provides ``special" locations on $Y$ which source Seiberg duality transformations. First let us see the magnitudes of the ``charges" (or the sizes of the steps) which come from the differences in the slopes in $e^{-\Phi}B_2$
given by (\ref{b2sol}) and (\ref{B2sol}) after and before a Seiberg duality transformation,
\ba
\Bigl(D_{l+1}-D_{l}
\Bigr)\frac{\pi \alpha'}{2}&=&\left(C_{l}+C_{l-2}-2C_{l-1}\right) \frac{g_s M \alpha'}{4} \nonumber \\
&=&\frac{3g_s M \alpha'}{4(K-l) (K-l+1) (K-l+2)}
 \label{fracN5a}
\ea
Note that, in the early stages of the duality cascade, the jump
is of order $\mathcal{O}(1/K^3)$ and is quite small for large $K$.
The magnitude increases as $l$ increases, and the
maximum value occurs at the last duality transformation where
$(C_{K-1}+C_{K-3}-2C_{K-2})=1/2$ for $N=KM$. If we sum up the charges at each\textbf{}
step, the magnitude of the total charge from the $K-1$ duality
transformations is proportional to
\be
\sum_{l=1}^{K-1}\left(C_{l}+C_{l-2}-2C_{l-1}\right)=
\frac{3}{4}\frac{(K-1) (K+2)}{K(K+1)}= \frac{3}{4}\Bigl(1-\frac{2}{K(K+1)}\Bigr).
\label{sumnsfq}
\ee
We see that in the limit $K\to \infty$, (\ref{sumnsfq}) goes to $3/4$. Of this the $(K-1)^{th}$ duality transformation takes $1/2$ and the $(K-2)^{th}$ duality transformation takes $1/8$. Presumably, these charges come from $NS5$-branes.
Most of the charge is located in the bottom region of the throat.
Moreover, the magnitude of the charges confirms that
the theory flows to a baryonic branch rather than to a confining branch and the duality cascade ends with the gauge group $SU(2M)\times SU(M)$, in the case where $N$ is an integral multiple of $M$. That
is because an additional duality transformation would need an infinite amount of charge, since
(\ref{fracN5a}) diverges for $l=K$. This implies either that the perturbative expansion of the anomalous dimension in powers of $M^{2}/N(N+M)$ is invalid here, or that the duality cascade stops at the baryonic branch. This latter case also agrees with the discussion and expectation in \cite{Gubser:2004qj} and \cite{Butti:2004pk} of a flow along a baryonic branch.
Recall that Seiberg duality in $\mathcal{N}=1$ supersymmetric $SU(N_c)$ gauge theory with $N_f$ flavors takes place in the so-called conformal window $3N_{c} > N_{f} > 3N_{c}/2$ and in the free magnetic phase $3N_{c}/2 \ge N_{f} >N_{c} +1$. To satisfy the last condition,
$K>2$. So we expect the last transition to take place at
$SU(3M) \times SU(2M) \rightarrow SU(M) \times SU(2M)$ and not go any further. This is consistent with our analysis.

The torsion and the flux components in the $6\oplus\bar{6}$ sector for the ansatz given in \cite{Papadopoulos:2000gj} and used in \cite{Butti:2004pk}
can be schematically written as $W_3^{(6)}=W_R+i W_I$, $H_3^{(6)}=H_R+i H_I$, and $F_3^{(6)}=F_R+i F_I$ and be split into two sets where the elements of a set have components in the same directions: $(W_R,\,F_I,\,\star_6 F_R,\,H_R,\,\star_6 H_I)$ and $(W_I,\,F_R,\,\star_6 F_I,\,H_I,\,\star_6 H_R)$ \cite{H2007}. In other words, the elements in the first set, $W_R,\,F_I,\cdots$ have components along some directions such as ${G_1}{ {\wedge}
   }{G_3}{ {\wedge} }{G_6}$ while the elements in the second set, $W_I,\,F_R,\cdots$ have components along other directions such as ${G_1}{ {\wedge}
   }{G_3}{ {\wedge} }{G_5}$.
If we let $\alpha$ and $\beta$ have arbitrary phase $\theta$ between them and write $\beta= \tan{\frac{w}{2}}e^{i\theta}\alpha$, then the equations of motion in the $6+\bar{6}$ sector, (\ref{W3F6})-(\ref{H6F6}), give
\be
\Bigl(1-\tan^2\frac{w}{2}\cos2\theta\,\Bigr)W_R =-2\tan\frac{w}{2}e^{\Phi}
\sin\theta\, F_I,
\ee
\be
\Bigl(1-\tan^2\frac{w}{2}\cos2\theta\,\Bigr)W_I =2\tan\frac{w}{2}e^{\Phi}
\sin\theta\, F_R,
\ee
\be
\tan^2\frac{w}{2}\sin2\theta\,W_I =2\tan\frac{w}{2}e^{\Phi}
\cos\theta\, F_R,
\ee
\be
-\tan^2\frac{w}{2}\sin2\theta\,W_R =2\tan\frac{w}{2}e^{\Phi}
\cos\theta\, F_I,
\ee
\be
\Bigl(1-\tan^2\frac{w}{2}\cos2\theta\,\Bigr)H_R =e^{\Phi}\Bigl(1+\tan^2\frac{w}{2}\cos2\theta\,\Bigr)
\star_6 F_R,
\ee
\be
\Bigl(1-\tan^2\frac{w}{2}\cos2\theta\,\Bigr)H_I =e^{\Phi}\Bigl(1+\tan^2\frac{w}{2}\cos2\theta\,\Bigr)
\star_6 F_I,
\ee
\be
\tan^2\frac{w}{2}\sin2\theta\, H_I=-e^{\Phi}\tan^2\frac{w}{2}\sin2\theta\, \star_6 F_I,
\ee
\be
-\tan^2\frac{w}{2}\sin2\theta\, H_R=e^{\Phi}\tan^2\frac{w}{2}\sin2\theta\, \star_6 F_R.\label{gen66ce}
\ee
We see that the equations of motion are over-constrained and do not have solution for generic values of $\theta$ unless $\theta=\pm\frac{\pi}{2}$ (or for $\theta=0,\pi$ with the fluxes and the torsions relabeled or related by S-duality).

In order to study the solutions which describe the effects of corrections to the anomalous mass dimension on the supergravity side, one needs a more general ansatz such as a complexification of the components $F_R$, $F_I$, $H_R$ and $H_I$ above and/or turning on $F_1$ flux in such a way that arbitrary phase between the two spinors could be accommodated. That is because the corrections change the imaginary self-duality condition in the 3-form combination $F_3-ie^{-\Phi}H_3$ in the KS solution in such a way that the supergravity flow does not occur at a fixed phase between the two spinors.
The change in magnitude of corrections after a Seiberg duality transformation leads to a step in the function $\tan{(w/2)}\,e^{i\theta}$ which relates the two spinors. This results in a geometric obstruction. These special geometric locations and the charges on them provide a gravitational source for Seiberg duality transformations.


\section{Discussions}

The gauge/gravity duality implies that the nonperturbative dynamics
of the gauge theory knows about the string background geometry. Here
we have studied the implications of corrections to the anomalous
mass dimension in the physical running of the couplings in the gauge
theory to the dual gravity theory in the KS throat. We find that a
more precise anomalous dimension on the gauge theory side reveals
structures on the gravity side. The corrections make the dilaton and
the potentials run with kinks and the fluxes and the metric have steps and
the deviation from KS grows more and more as the duality cascade
proceeds and the theory flows down to the bottom of the throat.

The magnitudes of the charges at the steps (or the sizes of the steps) are much smaller in the
early stages of the cascade than in the last steps.
The magnitudes of the charges confirm that
the theory flows to a baryonic branch rather than to a confining branch and the duality cascade ends with the gauge group $SU(2M)\times SU(M)$, in the case where $N$ is an integral multiple of $M$. That is because an additional duality transformation would require an infinite charge or a step with infinite size. This is consistent with what we would expect from the gauge theory side, since if we think of the $SU(M)$ in $SU(2M)\times SU(M)$ as a weakly gauged flavor symmetry, we have $SU(2M)$ with $2M$ flavors which falls far outside Seiberg's electric-magnetic duality window. Rather, it has a quantum deformed moduli space \cite{Seiberg:1994bz}. This also agrees with the discussion and expectation in \cite{Gubser:2004qj} and \cite{Butti:2004pk} of a flow along a baryonic branch.
Conversely, a duality cascade ending in the baryonic branch supports our premise that the anomalous dimension changes after every Seiberg duality transformation as the matter content of the theory changes with the $SU(2M)\times SU(M)$ not undergoing a duality transformation.

The steps also provide special locations with geometric obstructions which source Seiberg duality transformations. The steps we have discussed here are sharp because a Seiberg duality transformation occurs at a fixed $T^{1,1}$ base (or $AdS_5$ radius). The steps in the fluxes and the metric could be smoothed out if the charges at the steps get redistributed and a Seiberg duality transition takes place over some range of scales or  $AdS_5$ radius. Conversely, the sharpness of the steps on the gravity side would provide a measure for the sharpness of Seiberg duality transitions on the gauge theory side.
Overall, we expect that the relations (\ref{t1p2phi}) and (\ref{t1m2phi}), i.e., the gauge/gravity duality dictionary, should be modified by corrections. This is an important problem to be studied further.

It is believed that $NS5$-brane charges are located at the bottom of the KS throat. The $NS5$-branes wrap an $S^{2}$ of the $S^{3}$. In the absence of fluxes, they wrap a shrinking $S^{2}$ of the $S^{3}$ at, say, angular coordinate $\psi=0$ and essentially vanish without a trace. If such $NS5$-brane can tunnel to
 the other pole of the $S^{3}$,{\it i.e.}, $\psi=\pi$, it becomes $M$ $D3$-branes  \cite{Kachru:2002gs}.
Since there are $K$ $NS5$-branes, together there are $N=KM$ $D3$ charges. Our picture suggests
that these $K$ $NS5$-branes are located at different positions, one at each $r=r_{l}$. Furthermore, they are at different locations of the $S^{3}$, that is, they wrap different shrinking $S^{2}$s of the $S^{3}$.
As a consequence, $H_{3}$ is in general no longer orthogonal to $F_{3}$, which is the case with a non-zero R-R 1-form flux $F_{1}$.

In this paper, we work with the ansatz that, for large $r$, ${\bf H}_{3}$ is along the direction of
$dr \wedge \omega_{2}$. Although we need only the magnitude of $F_{3}$ (as given by (\ref{hgensola}) and (\ref{hgensolb})) to find the warp factor, it is important to find the components of ${\bf F}_{3}$ (since ${\bf F}_{3}$ does not all lie in $\omega_{3}$) and an explicit solution of the supergravity equations, even if only perturbatively in $1/K$.

The gauge/gravity duality is a powerful tool to probe both the gauge and the
dual gravity theories from different directions.
It seems that the best way to test/use gauge/gravity duality
is an attempt to find the gravity dual to QCD. However, our
theoretical control is best when we consider super Yang-Mills
theories. Unfortunately, lattice gauge theory for super Yang-Mills
theories is too rudimentary to be useful at the moment. Since
Seiberg duality  and gauge/gravity duality are strongly believed but
not proven, it is quite amazing and useful that the whole notion of
Seiberg duality and gauge/gravity dualities may be tested in cosmology. If
our universe indeed resides at a bottom of such a warped deformed
throat, the cosmological implication of this step-wise (or cascading) behavior of
the metric on the brane inflationary scenario and the cosmic
microwave background radiation can be very interesting.

\section*{Acknowledgements}

We thank Xingang Chen, Igor Klebanov, Oleg Lunin, Sarah Shandera and Jiajun Xu for discussions.
This research is supported in part by the National Science Foundation under grant
number NSF-PHY/03-55005. G.H. is a Provost's Fellow.

\bibliographystyle{JHEP}

\providecommand{\href}[2]{#2}\begingroup\raggedright\endgroup

\end{document}